\documentclass[twocolumn,10pt,amsmath,amssymb,superscriptaddress,pra,aps]{revtex4-1}
\usepackage{epsfig}
\usepackage{color}
\usepackage{graphicx}
\usepackage{bm}
\usepackage{epstopdf}
\usepackage[T1]{fontenc}
\usepackage{hyperref}
\hypersetup{
  colorlinks   = true,  
  urlcolor     = black,  
  linkcolor    = blue,  
  citecolor    = red    
} 
 
\begin{document} 
\title{Intercomponent correlations \\ in attractive one-dimensional mass-imbalanced few-body mixtures}
\author{Daniel \surname{P{\k e}cak}}
\affiliation{\mbox{Institute of Physics, Polish Academy of Sciences, Aleja Lotnikow 32/46, PL-02668 Warsaw, Poland}} 
\author{Tomasz \surname{Sowi\'{n}ski}}
\affiliation{\mbox{Institute of Physics, Polish Academy of Sciences, Aleja Lotnikow 32/46, PL-02668 Warsaw, Poland}}
\date{\today}
\begin{abstract}
Ground-state properties of a few attractively interacting ultra-cold atoms of different mass confined in a one-dimensional harmonic trap are studied. 
The analysis is performed in terms of the noise correlation, which captures the two-particle correlations induced by the mutual interactions. 
Depending on the mass ratio between the components' atoms, the inter-particle correlations change their properties significantly from a strong pair-like correlation to an almost uncorrelated phase. This change is accompanied by a simultaneous change in the structure of the many-body ground state. A crucial role of the quantum statistics is emphasized by comparing properties of the Fermi-Fermi mixture with a corresponding Fermi-Bose system.
\end{abstract}
\maketitle

\section{Introduction}
Recent years have brought many examples that the systems consisting of a few ultra-cold atoms can be prepared and well-controlled experimentally with extreme precision \cite{wenz2013fewToMany,zurn2012fermionization,serwane2011deterministic,Murmann2015AntiferroSpinChain,Kaufman2015Entangling,Andrea2018Imaging,zurn2013Pairing}. 
It became possible to measure (as functions of mutual interactions) not only single-particle properties of the system but also higher multi-particle correlations. The latter are fundamentally important since they directly reflect different non-classical multi-particle properties of the system being direct manifestations of indistinguishability and entanglement forced by interactions.
It is quite obvious that these higher correlations cannot be neglected if one needs to characterize an obtained quantum state appropriately. As shown recently, the two-body position and momentum correlation functions can be measured and they are indeed a very powerful tool to characterize quantum states \cite{bergschneider2018correlations}. In principle, one can have experimental access also to higher-order correlations between particles. For example, it can be done by using atomic microscopes which allow one to measure positions of all particles at the same time~\cite{bakr2010probing,sherson2010single,OmranBloch2015PauliBlocking, Cheuk2015FermionicMicroscope,Cheuk2016MottMicroscope,Cheuk2016Correlations, ParsonsPRL2015,EdgePRA2015ImagingFermions,haller2015single,ParsonsPRL2015}. All this means that on an experimental level, the ultra-cold physics starts to explore much more complicated features of many-body systems than simple single-particle densities.   

Theoretical studies of one-dimensional few-body mixtures are very rich in the literature. In a great majority, due to the experimental motivation from the Heidelberg group \cite{serwane2011deterministic,zurn2013Pairing}, these considerations concern two-component mixtures of repelling particles with equal mass (see for example \cite{brouzos2013two,SowinskiGrass2013FewInteracting,Gharashi2013UpperBranchCorrelations,GarciaMarch2014Localization,fogarty2018fast,Garcia-MarchPRA2014,Valiente2014Multi,volosniev2015hyperspherical,koscik2012quantum}). Accordingly less attention is given for attractive systems for which some precursors of the Cooper-like pairing were observed \cite{zurn2013Pairing}, theoretically explained \cite{zinner2013comparing,DAmico2015Pairing,Sowinski2015Pairing} and explored~\cite{conduit2013fflo,rammelmuller2016ground,mckenney2016ground,BjerlinReimann2016Higgs}. 

Recent years have brought tremendous progress in experimental studies of fermionic (Li-K, Dy-K)~\cite{Wille6Li40K,tiecke2010Feshbach6Li40K,cetina2016ultrafast,Grimm2018DyK} as well as bosonic-fermionic (Li-Na, Rb-K, Cs-Li, Li-K) \cite{Hadzibabic2002LiNa,Esslinger2006RbK,Bloch2009RbK,Wu2011KKLi,Chin2014CsLi,Grimm2018KLi} mixtures of large number of ultra-cold atoms of different mass.
Although for such mixtures the few-body regime has not been achieved yet, first theoretical predictions show that such systems may have essentially different properties \cite{JasonHo2013PhaseSeparation,loft2014variational,Pecak2016Separation,Pecak2017Ansatz,harshman2017masses,mistakidis2018repulsive} than systems with equal mass atoms.

In this general context, the question of properties of mass-imbalanced few-body mixtures in the attractive regime seems to be very relevant and important. In the following, we perform the first step in this direction and we analyze a destructive effect of a mass difference on inter-component pairing correlations emerging in a strong attractive regime. We identify, describe, and quantify these highly non-classical correlations as functions of mass ratio between particles forming opposite components and their number. We also emphasize the role of the quantum statistics in this destructive process.

The work is organized as follows. In Sec.~\ref{sec:model} we introduce the theoretical model of the few-body ultra-cold system studied and we briefly discuss a numerical method of treatment used.  Next in Sec.~\ref{sec:noise}, we refresh the concept of noise correlation and we introduce a natural measure quantifying an amount of inter-component correlations in the system. In Sec.~\ref{sec:twoatoms} we analyze the simplest situation of two atoms for repulsive and attractive interactions and different masses. Importantly, in Sec.~\ref{sec:manybody} we broadly discuss inter-component correlations induced by attractions for different strengths of interactions and different mass ratios. For completeness, in Sec.~\ref{sec:boseFermi} we examine the consequences of the quantum statistics by studying Bose-Fermi mixtures. Finally, in Sec.~\ref{sec:conclusion} we conclude.

\section{The Model} \label{sec:model}
In the following we consider a two-component mixture of ultra-cold fermions of masses $m_\downarrow$ and $m_\uparrow$ confined in a one-dimensional harmonic trap of a frequency $\omega$. We assume that the particles belonging to different species interact dominantly in the $s$-wave channel and we model this interaction with the $\delta$-like potential. In contrast, for fermions of the same kind (for which the $s$-wave channel is closed due to the Pauli exclusion principle) mutual interactions are negligible and we ignore them. Under these assumptions the many-body Hamiltonian of the system reads:
\begin{multline} \label{eq:ham}
\hat{\cal H} = \sum_{\sigma\in\{\uparrow,\downarrow\}} \int\!\!\mathrm{d}x\,\hat{\Psi}^\dag_\sigma(x)H_\sigma\hat{\Psi}_\sigma(x)  \\
 + g\int\!\! \mathrm{d}x\,\hat{\Psi}^\dag_\uparrow(x)\hat{\Psi}^\dag_\downarrow(x)\hat{\Psi}_\downarrow(x)\hat{\Psi}_\uparrow(x),
\end{multline}
where the single-particle Hamiltonians $H_\sigma$ are given by:
\begin{subequations} \label{1PHam}
\begin{align}
 H_\downarrow &= - \frac{\hbar^2}{2m_\downarrow} \frac{\mathrm{d}^2}{\mathrm{d} x^2} + \frac{m_\downarrow\omega^2}{2} x^2, \\
 H_\uparrow   &= - \frac{\hbar^2}{2m_\uparrow} \frac{\mathrm{d}^2} {\mathrm{d} x^2} + \frac{m_\uparrow\omega^2}{2} x^2.
\end{align}
\end{subequations}
Note, that for simplicity the frequencies $\omega$ are equal for both components and that implies only one energy scale $\hbar\omega$ in the system.
The field operator $\hat\Psi_\sigma(x)$ annihilates a particle of type $\sigma$ at given point $x$. The quantum statistics is reflected in the natural anti-commutation relations, $\{\hat\Psi_\sigma(x),\hat\Psi^\dagger_\sigma(x')\} = \delta(x-x')$ and $\{\hat\Psi_\sigma(x),\hat\Psi_\sigma(x')\} = 0$. Note that particles of different types are fundamentally distinguishable. Therefore any appropriate relations between fields $\hat\Psi_\uparrow(x)$ and $\hat\Psi_\downarrow(x)$ are equivalent to the commutation relations $[\hat\Psi_\uparrow(x),\hat\Psi^\dagger_\downarrow(x')] = [\hat\Psi_\uparrow(x),\hat\Psi_\downarrow(x')]=0$. Evidently, the Hamiltonian \eqref{eq:ham} does commute with the particle number operators in a given component $\hat{N}_\sigma=\int\mathrm{d}x\hat{\Psi}_\sigma^\dagger(x)\hat{\Psi}_\sigma(x)$. Therefore, the properties of the system can be examined independently in the subspaces of given $N_\uparrow$ and $N_\downarrow$. To make a whole analysis as clear as possible, in this work we focus on balanced systems, {\it i.e.}, the systems with equal number of particles in both components, $N_\uparrow=N_\downarrow$. The effective one-dimensional interaction strength $g$ between fermions from opposite components can be derived from the full three-dimensional theory of scattering by integrating out the perpendicular degrees of motion \cite{Olshanii1998}. In the following we express all quantities in the natural harmonic oscillator units with respect to the $\downarrow$ component, {\it i.e.}, energies are measured in $\hbar\omega$, positions in $\sqrt{\hbar/(m_\downarrow\omega)}$, the interaction strength $g$ in units of $(\hbar^3\omega/m_\downarrow)^{1/2}$, {\it  etc.} For convenience we also denote the mass ratio of atoms of different species as $\mu=m_\uparrow/m_\downarrow$.   

By expanding the field operators $\hat{\Psi}_\sigma(x) = \sum_i \phi_{i\sigma}(x) \hat{a}_{i\sigma}$ in the eigenbasis $\{\phi_{i\sigma}(x)\}$ of the appropriate single-particle Hamiltonians \eqref{1PHam} we rewrite the Hamiltonian \eqref{eq:ham} to the following form:
\begin{equation} \label{eq:hamSQ}
 \hat{{\cal H}} = \sum_\sigma \sum_i E_i \hat{a}_{i\sigma}^\dag \hat{a}_{i\sigma} + 
 g \sum_{ijkl} U_{ijkl}  \hat{a}_{i\downarrow}^\dag \hat{a}_{j\uparrow}^\dag \hat{a}_{k\uparrow}^{\phantom{\dag}} \hat{a}_{l\downarrow}^{\phantom{\dag}},
\end{equation}
where $E_i$ are the single-particle eigenenergies and the interaction coefficients $U_{ijkl}$ read:
\begin{equation} \label{eq:uijkl}
U_{ijkl} = \int \mathrm{d}x \phi^*_{i\downarrow}(x) \phi^*_{j\uparrow}(x) \phi_{k\uparrow}(x) \phi_{l\downarrow}(x).
\end{equation}

To find numerically the ground-state $|G_0\rangle$ of an interacting system, we calculate the matrix elements of the Hamiltonian~\eqref{eq:hamSQ} in the Fock space spanned by the many-body noninteracting Fock states $\{|F_j\rangle\}$ constructed from the lowest single-particle orbitals and we diagonalize the matrix obtained via the Arnoldi method \cite{ARPACK1998Sorensen}. In this way we find the decomposition of the many-body ground-state in this basis, $|G_0\rangle = \sum_j \alpha_j|F_j\rangle$. The size of the cropped Fock space is carefully selected in such a way that the final results are almost insensitive for further extension of the Fock space. 
It is worth to note that the many-body eigenenergies obtained via the exact diagonalization method converge slowly with the size of the Hilbert space. However, in the case of the many-body wave functions the convergence, understood as a numerical stabilization of coefficients in the decomposition of the many-body ground-state into the Fock basis, is much faster (for details see for example \cite{Pecak2017com} and Appendix~\ref{AppendixConv}). Since in the following article we focus on the two-body correlations, therefore a poor convergence of the ground-state energy does not reduce the credibility of the results.
We checked that in the cases studied it is sufficient to use the first ten single-particle states in the decomposition of the field operator $\hat{\Psi}_\sigma(x)$ to obtain reliable results.
The corresponding sizes of the many-body Hilbert space is presented in Tab.~\ref{tab}.
\begin{table}
\begin{center}
\begin{tabular}{l||r|r}
 & \hspace{3mm}{\bf Fermi-Fermi}\hspace{0mm} & \hspace{3mm}{\bf Bose-Fermi}\hspace{0mm} \\ 
 \hline \hline
$N_\downarrow=N_\uparrow=2$ &  2 025 &  2 475\\
\hline 
$N_\downarrow=N_\uparrow=3$ & 14 400 & 26 400\\
\hline
$N_\downarrow=N_\uparrow=4$ & 44 100 & 150 150\\
\hline
$N_\downarrow=N_\uparrow=5$ & 63 504 & 504 504
\end{tabular}
\end{center}
\caption{\label{tab} 
The size of the cropped many-body Hilbert space for different number of particles occupying no more than ten first single-particle orbitals $\phi_{i\sigma}(x)$, $i\in\{0,\ldots 9\}$.
Note, that for Bose-Fermi mixtures the size of the Hilbert space is significantly larger.}
\end{table}

\section{Noise correlation}\label{sec:noise}
The simplest observable that characterizes an interacting few-body system is the single-particle density profile being the diagonal part of the single-particle density matrix:
\begin{equation} \label{eq:SPx}
 \rho_\sigma^{(1)}(x) = \langle G_0 | \hat{\Psi}^\dag_\sigma (x) \hat{\Psi}_\sigma (x) | G_0 \rangle.
\end{equation}
It can be simply understood as the probability density of finding a single particle from the component $\sigma$ at position~$x$. Similarly, the probability density of finding a single particle with the momentum $p$ reads:
\begin{equation} \label{eq:SPp}
 \tau_\sigma^{(1)}(p) = \langle G_0 | \hat{\Psi}^\dag_\sigma (p) \hat{\Psi}_\sigma (p) | G_0 \rangle.
\end{equation}
Note that in the latter definition the field operator is expressed in the momentum domain. These two single-particle quantities are the simplest (apart from the energy of the state) measurable observables which characterize the many-body quantum system. However, they do not possess any information about correlations between simultaneously measured particles belonging to opposite components. These features are captured by the two-body correlations which are encoded complementarily in the two-particle densities in position and momentum domains:
\begin{subequations}\label{eq:TP}
\begin{equation} \label{eq:TPx}
 \rho^{(2)}(x;y) = \langle G_0 | \hat{\Psi}^\dag_\downarrow(x)  \hat{\Psi}^\dag_\uparrow(y) \hat{\Psi}_\uparrow(y) \hat{\Psi}_\downarrow(x) | G_0 \rangle,
\end{equation}
\begin{equation} \label{eq:TPp}
 \tau^{(2)}(p;k) = \langle G_0 | \hat{\Psi}^\dag_\downarrow(p) \hat{\Psi}^\dag_\uparrow(k) \hat{\Psi}_\uparrow(k) \hat{\Psi}_\downarrow(p)  | G_0 \rangle.
\end{equation}
\end{subequations}
In the noninteracting case ($g=0$) the many-body wave function of the ground state is a simple product of two antisymmetric wave functions, one for each component.
Consequently, two-particle densities are the products of the corresponding single-particle densities, $\rho^{(2)}(x;y) = \rho^{(1)}_\downarrow(x) \rho^{(1)}_\uparrow(y)$ and $\tau^{(2)}(p;k) = \tau^{(1)}_\downarrow(p) \tau^{(1)}_\uparrow(k)$. When the interactions are turned on ($g\neq 0$), these relations do not hold anymore since inter-component correlations emerge in the system. It turns out that these additional correlations forced by interactions are well captured by the so-called noise correlations introduced in \cite{lukin2004noisCorr,Altman20081D,Altman2009lowDim} and exploited recently in the context of repulsive few-body systems~\cite{brandt2017two,brandt2018HOM}. These quantities are defined as the following:
\begin{subequations}\label{eq:Noise}
\begin{equation} \label{eq:NoiseX}
  {\cal G}_\rho(x;y) = \rho^{(2)}(x;y) - \rho^{(1)}_\downarrow(x) \rho^{(1)}_\uparrow(y),
\end{equation}
\begin{equation} \label{eq:NoiseP}
 {\cal G}_\tau(p;k) = \tau^{(2)}(p;k) - \tau^{(1)}_\downarrow(p) \tau^{(1)}_\uparrow(k)
\end{equation}
\end{subequations}
and they just show the differences between the exact two-particle densities and one predicted by the single-particle picture.
It is worth noting that the noise correlation can be measured experimentally in the position as well as in the momentum domain~\cite{folling2005spatial,bergschneider2018correlations}.

As explained above, the noise correlations \eqref{eq:Noise} are appropriate quantifiers of inter-component correlations. However, having two different noise correlations for two different experimental parameters it is very hard to select one having higher correlations. Therefore, it is very convenient to introduce some geometric distance between an actual two-particle density profile and that obtained as a product of single-particle ones. Fortunately, single- and two-particle density profiles have all mathematical properties of density distributions. Therefore, the natural metric in their space exists. The Frobenius distance (known also as Hilbert-Schmidt norm) \cite{deza2009encyclopedia,rachev2013methods} can be extracted directly from the noise correlations:
\begin{subequations}\label{eq:measure}
\begin{align}
 ||{\cal G}_\rho|| &= \left(\int\!\mathrm{d}x\,\mathrm{d}y\, |{\cal G}_\rho(x;y)|^2\right)^{1/2}, \\
 ||{\cal G}_\tau|| &= \left(\int\!\mathrm{d}p\,\mathrm{d}k\, |{\cal G}_\tau(p;k)|^2\right)^{1/2}.
\end{align}
\end{subequations}
It is quite obvious that the distance vanishes for the noninteracting system and it grows when an average magnitude of the inter-component correlations increases. In the following, we will quantify correlations mainly in the language of this quantity. 

\section{Two-atom system}\label{sec:twoatoms}
\begin{figure}[t]
\centering
\includegraphics[width=\columnwidth]{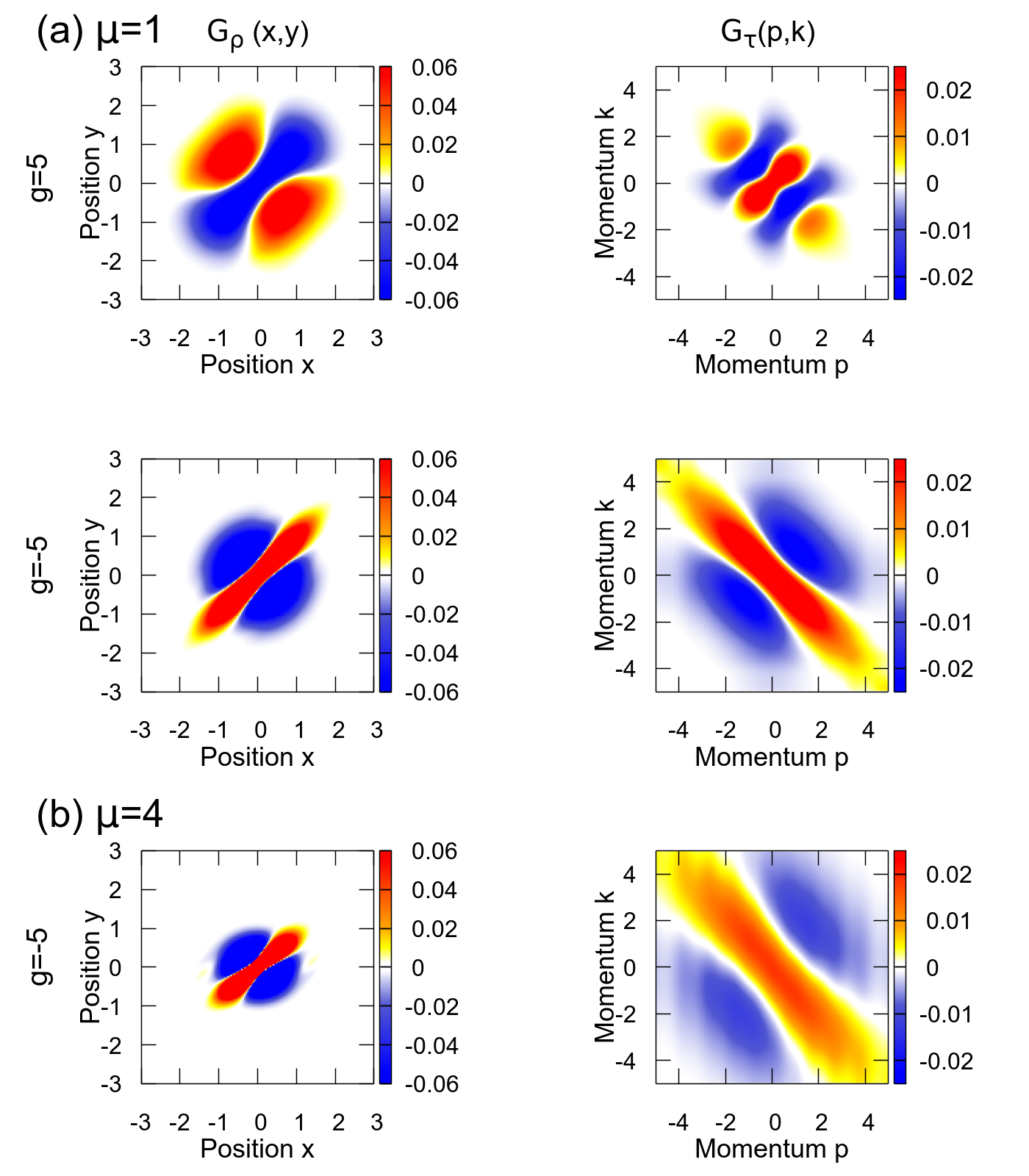}
\caption{The noise correlations ${\cal G}_\rho$ and ${\cal G}_\tau$ for a system of two particles ($N_\uparrow=N_\downarrow=1$). (a) The equal mass case ($\mu=1$) for strong inter-particle repulsions (upper row) and attractions (bottom row). 
(b) Strongly mass-imbalanced case ($\mu=4$) for strong inter-particle attractions. Note, that independently of the mass ratio the noise correlations demonstrate similar correlations (anti-correlations) in the position (momentum) space. See the main text for explanation.
In all plots positions and momenta are measured in natural units of the harmonic oscillator, $\sqrt{\hbar/m_\downarrow\omega}$ and $\sqrt{\hbar m_\downarrow\omega}$, respectively.
}
\label{fig1}
\end{figure}
Before we analyze inter-component correlations for a larger number of attractively interacting particles let us start from the simplest situation of two particles of equal mass ($\mu=1$) for which the exact analytical expression for the ground-state wave function and its energy is known~\cite{Busch1998}. In the context of the noise correlation, the ground state of a few repelling fermions ($g>0$) was considered recently in \cite{brandt2017two}, where the interactions were modeled by the Gaussian-shaped inter-particle potential. The width of the Gaussian was much smaller than the natural harmonic oscillator length, hence we reproduce the results by using the pure $\delta$ potential.
As seen in the upper row of Fig.~\ref{fig1}a, for strong repulsions ($g=5$) the noise correlation ${\cal G}_\rho$ become negative on the diagonal. As noticed in \cite{brandt2017two}, it is a direct manifestation of the fact that due to repulsions it is almost not possible to find two particles in the same position. Importantly, this effect cannot be captured by a simple product of single-particle densities. For the same interaction, also some non-trivial behavior of the noise correlation in the momentum domain ${\cal G}_\tau$ is present (right panel in Fig.~\ref{fig1}a). 

The situation changes qualitatively for the attractive scenario ($g=-5$). In this case, the probability of finding two particles at the same position is highly enhanced when compared to quite poor predictions of single-particle distributions. The most prominent difference between repulsive and attractive systems is however visible in the momentum domain. For an attractive system, one finds a very strong anti-correlation between interacting particles signified as a high positive value of the noise correlation along the line $p=-k$ (see the bottom right plot in Fig.~\ref{fig1}a). This means that the probability of finding two particles having exactly opposite momenta is significantly larger than that predicted by the single-particle picture. 
Importantly, in the case of the two particles studied, the situation does not change significantly when different masses of particles are considered. As it is seen in Fig.~\ref{fig1}b, even for a large mass ratio ($\mu=4$), 
the strong correlation in positions and anticorrelation in momenta are present in the system.

\section{Many-body system} \label{sec:manybody}
\begin{figure}[t]
\centering
\includegraphics[width=\columnwidth]{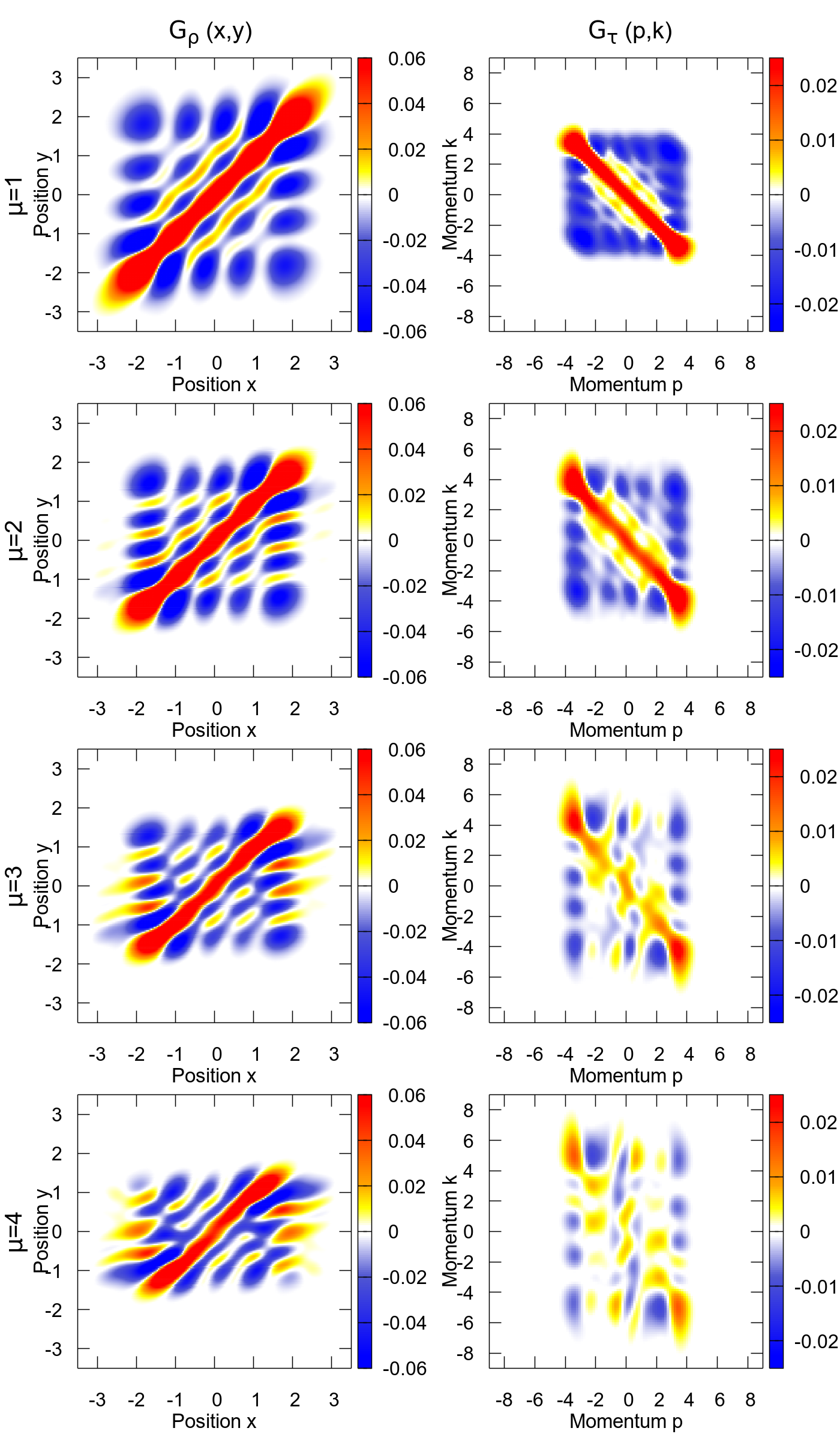}
\caption{The noise correlations ${\cal G}_\rho$ and ${\cal G}_\tau$ for a strongly attractive system ($g=-5$) of $N_\uparrow=N_\downarrow=4$ particles and different mass ratio $\mu=1,2,3,4$. Note that while the strong correlations in the position domain survive when the mass imbalance in the system is introduced, the anti-correlation in the momentum domain is strongly suppressed and reduced. Position and momentum are measured in the natural units of the harmonic oscillator, $\sqrt{\hbar/m_\downarrow\omega}$ and $\sqrt{\hbar m_\downarrow\omega}$, respectively.}
\label{fig2}
\end{figure}
The two-particle system described above is trivial from the quantum statistics point of view. Therefore in this section, we focus on attractive systems with a larger number of particles ($N_\uparrow=N_\downarrow=4$). First, we calculate the noise correlation for the balanced system of equal-mass particles (upper row in Fig.~\ref{fig2}). As it is seen, the inter-component correlations for the attractive scenario ($g=-5$) qualitatively resemble main features observed in the two-body scenario -- strong correlations in positions and anti-correlations in momenta are clearly visible. It should be noted, however, that the noise distribution in the momentum domain is much flatter along the line $p=-k$ than the corresponding one obtained for a smaller number of particles. This effect is forced by an inherent indistinguishability of fermions and it can be viewed as one of the indicators of the Cooper-like pairing in the system \cite{Sowinski2015Pairing}.
  
The situation changes when some factor lifting the balance in the system is present. In principle, in our case, there are two distinct mechanisms leading to the imbalance. The first originates in a direct difference of the number of particles in each component. The second is forced by different masses of the atoms forming opposite components ($\mu\neq 1$). In the case of harmonic confinement there exists a quite important difference between these two scenarios. It is clearly visible in the noninteracting limit. When a difference in numbers of particles is considered, contributions to the total energy of the system of both components are different (they have different Fermi energies). In contrast, the Fermi energy is insensitive to any change in the mass of particles since in the case of harmonic confinement the single-particle energies $E_i$ do not depend on mass (due to the same frequency $\omega$). All this suggests that these two different mechanisms may have a different impact on the properties of the system. In this work, we focus only on the imbalance forced by the mass difference ($\mu\neq 1$) assuming always a balance in the particle number $N_\uparrow=N_\downarrow$. 

As explained in previous works \cite{Pecak2016Separation,Pecak2016Transition}, with varying $\mu$ the single-particle harmonic orbitals change their shape and they become different for different components. Although the single-particle energies remain unchanged, the mutual repulsions force the system to excite lighter particles. As a consequence, for the repulsion ($g>0$) strong enough the separation of the density profiles emerge. 
A similar effect of the phase separation driven by the mass imbalance was also studied in the case of homogeneous systems and non-harmonic confinements~\cite{conduit2011itinerant,JasonHo2013PhaseSeparation,Fratini2014ZeroTemp}.

In the case of attractive interactions ($g<0$), the situation also changes when compared to the balanced system $\mu=1$. It is clearly visible when the noise correlations are considered (Fig.~\ref{fig2}). Although in the position domain the main effect caused by $\mu\neq 1$ is quite trivial (the distribution of the heavier component is just narrower), in the momentum domain the change is significant. As it is seen in Fig.~\ref{fig2}, the strong anti-correlations along the line $p=-k$ are smeared and for a large enough mass ratio $\mu$ any traces of correlated pairs almost vanish.
\begin{figure}[t]
\centering
\includegraphics[width=\columnwidth]{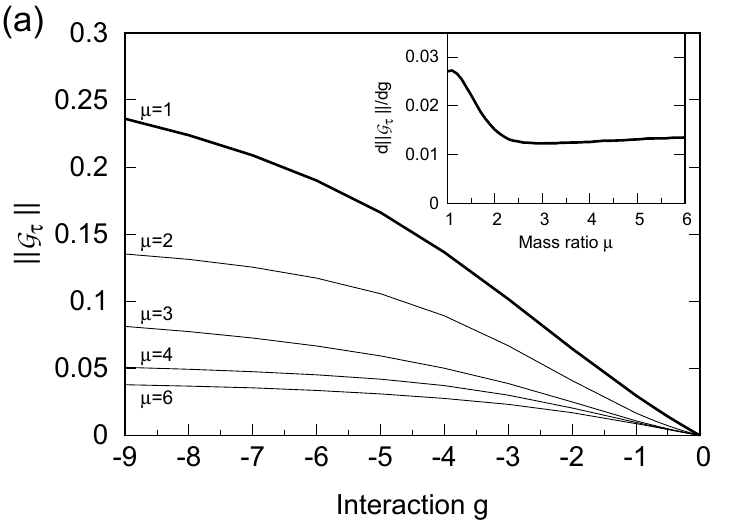}
\includegraphics[width=\columnwidth]{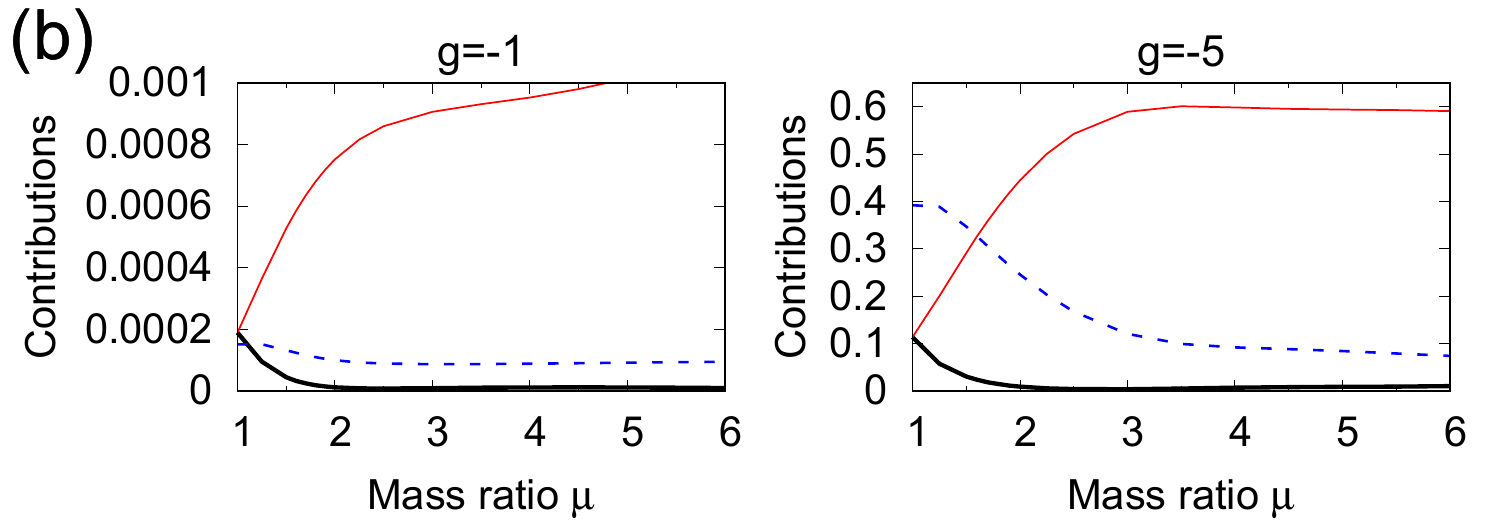}
\caption{(a) The total Frobenius distance $||{\cal G}_\tau||$ as a function of the interaction strength $g$ for the system of $N_\uparrow=N_\downarrow=4$ particles and different mass ratio $\mu$. In the inset we show the slope $\mathrm{d}||{\cal G}_\tau||/\mathrm{d}\mu$ calculated in the region of vanishing interactions~$g$. (b) Contributions ${\cal C}_1$ (solid thin), ${\cal C}_2$ (solid thick), and ${\cal C}_3$ (dashed) of different sectors of the many-body Hilbert space to the ground state of the system as a function of mass ratio $\mu$ for weak and strong interaction $g$. Note that around $\mu\approx 2$ (the exact value depends on $g$) two sectors are strongly suppressed and the excitations are dominated only by one type of states with $k=1$. The interactions are measured in the natural units of the harmonic oscillator, $(\hbar^3\omega/m_\downarrow)^{1/2}$. }
\label{fig3} 
\end{figure}
To show quantitatively how the inter-component correlations change with varying parameters of the system, first, we focus on the Frobenius distance $||{\cal G}_\tau||$ as a function of the attraction $g$ for the fixed mass ratio $\mu$ (Fig.~\ref{fig3}). From this figure, one can easily deduce the general behavior of the system. First, it is clearly visible that independently of the mass ratio $\mu$ the inter-component correlations grow with an amplitude of interactions. This fact is in full accordance with our intuition -- stronger inter-component forces lead to stronger correlations between particles. One also notices that for fixed interactions and increasing mass ratio $\mu$ correlations measured by $||{\cal G}_\tau||$ decreases, {\it i.e.}, particles become less correlated. We can quantify this behavior more precisely in a few different ways. The simplest one is by calculating the derivative $\mathrm{d}||{\cal G}_\tau||/\mathrm{d}g$ close to the perturbative regime ($-0.5\lesssim g<0$) where a linear growth of correlations is visible. As shown in the inset of Fig.~\ref{fig3}, in this range of interactions, the slope of the derivative (the second derivative of $||{\cal G}_\tau||$) evidently depends on mass and around $\mu\gtrsim 2$ the rate $\mathrm{d}||{\cal G}_\tau||/\mathrm{d}g$ becomes almost independent of $\mu$. To find the origin of this surprising change of the slope, we performed a direct numerical inspection of the many-body ground state. We find that in the case studied, around $\mu\approx 2$, a specific change of the ground-state structure is clearly visible. It can be viewed by performing a specific decomposition of the many-body ground state. Generally, the ground-state of the system can be written as a superposition of Fock states belonging to four disconnected sectors of the many-body Hilbert space:
\begin{equation}
|G_0\rangle = \alpha_0|F_0\rangle + \sum_{k=1}^{3}\sum_j \alpha^{(k)}_j |F^{(k)}_j\rangle.
\end{equation}
The first sector contains only the ground state of the noninteracting system $|F_0\rangle$. Three other sectors spanned by vectors $\{|F^{(k)}_j\rangle\}$ have the following properties. For $k=1$, the states $|F^{(1)}_j\rangle$ are products of the noninteracting ground state of $\uparrow$ (heavier) particles and different excited states of $\downarrow$ (lighter) ones. Conversely, for $k=2$, the states $|F^{(2)}_j\rangle$ are products of different excited states of $\uparrow$ (heavier) particles and the noninteracting ground state of $\downarrow$ (lighter) particles. Finally, for $k=3$ the states $|F^{(3)}_j\rangle$ are built only from the excited states in both components. Having this decomposition, one can calculate contributions from different sectors to the interacting ground state of the system. These contributions are quantified by four numbers ${\cal C}_k = \sum_j|\alpha_j^{(k)}|^2$ and ${\cal C}_0=|\alpha_0|^2$. Inter-component correlations are encoded in excitations of the system and therefore they are directly reflected in non-vanishing values of ${\cal C}_k$ with $k=1,2,3$. In Fig.~\ref{fig3}b we plot these quantities as functions of the mass ratio $\mu$ for $g=-1$ and $g=-5$. As it is seen, for equal mass system $\mu=1$, all sectors of the system's excitations contribute to building the correlations. However, when the mass ratio increases, one of the sectors ($k=1$) starts to dominate. At the same time the other sectors are strongly suppressed and from around $\mu_c\approx 2$ (the exact value depends on interaction strength) the many-body ground state can be written almost perfectly as a superposition of Fock states having all heavy fermions located in the lowest harmonic oscillator orbitals. It means that for $\mu>\mu_c$ the inter-component correlations are much less sensitive to any further variations of the mass ratio and they come only from variations of the internal structure of the lighter component.

\begin{figure}[t]
\centering
\includegraphics[width=\columnwidth]{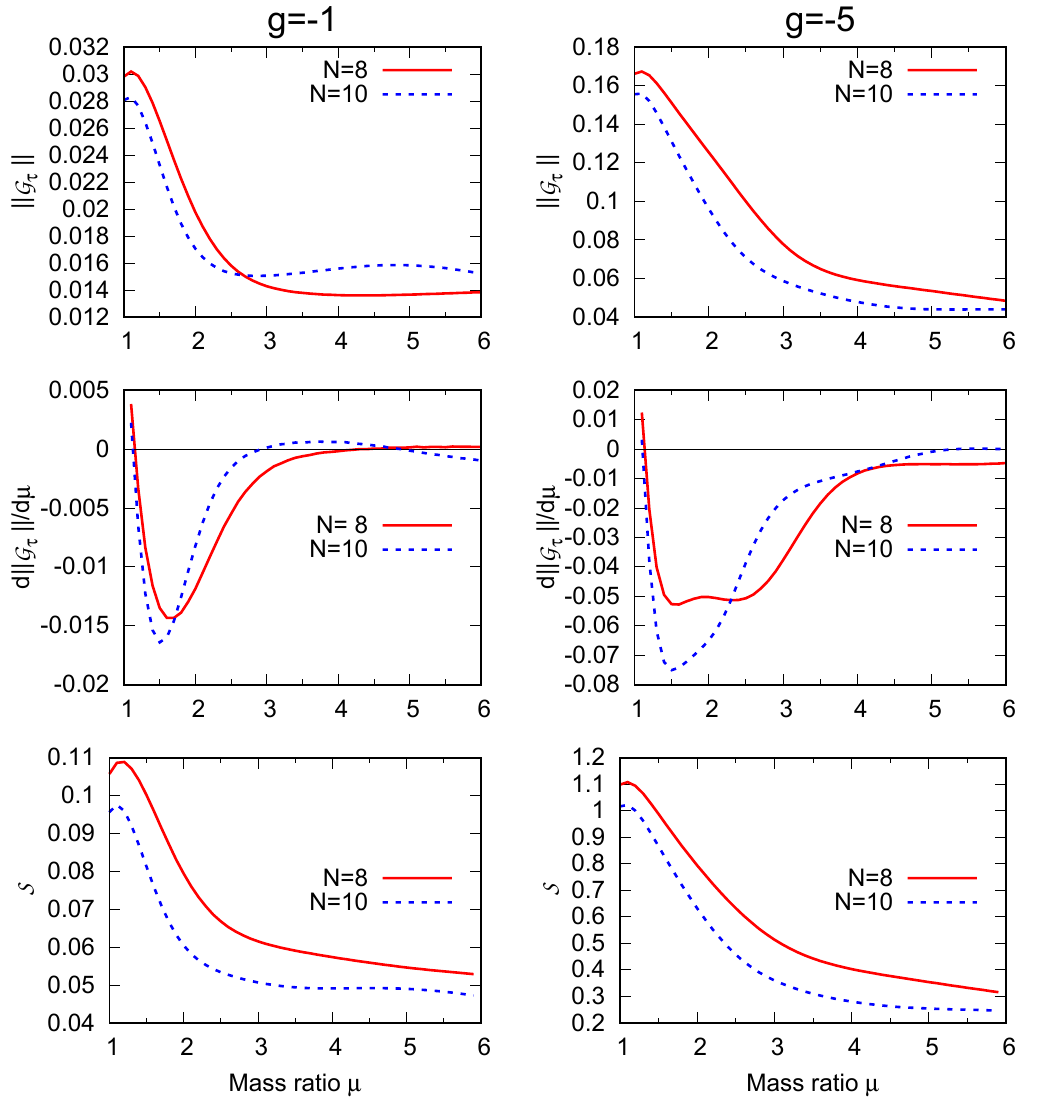}
\caption{The Frobenius distance $||{\cal G}_\tau||$ (upper row), its derivative $\mathrm{d}||{\cal G}_\tau||/\mathrm{d}\mu$ (middle row), and the von Neumann entropy $\cal S$ (bottom row) as functions of the mass ratio $\mu$ for different number of particles and different interactions. Note the clearly visible minima of derivatives at which the distances $||{\cal G}_\tau||$  and the entropies $\cal S$ significantly change their behaviour. It is related to the change in the structure of the many-body ground state of the system. See the main text for details.
}\label{fig4}
\end{figure}

The transition in the ground-state structure at around $\mu_c$ can be also visualized by plotting the distance $||{\cal G}_\tau||$ and its derivative $\mathrm{d}||{\cal G}_\tau||/\mathrm{d}\mu$ as functions of the mass ratio $\mu$ for different numbers of particles and different interactions (see Fig.~\ref{fig4}). As it is seen, for some particular value of the mass ratio $\mu_c\approx 2$ the derivative has a clearly visible minimum. Although the critical value $\mu_c$ depends on system parameters (interaction $g$, number of particles), the mechanism is always the same --- for mass ratio larger than $\mu_c\approx 2$ the exact form of the ground-state is significantly simplified and it manifests a very high probability of finding all heavy particles in their noninteracting ground state. 

To lower the energy of the attractive system it is preferred that the two kinds of particles have the same spatial distributions --- then the interaction integrals $U_{ijkl}$ are the largest. In principle, spatial distributions can be adjusted by exciting particles to higher single-particles orbitals. Although the excitation cost is the same for both components, due to the different length scales for the components, adjusting the density profile of heavier particles requires much more excitations. Therefore it is energetically favorable to excite light particles keeping the heavy component almost in the noninteracting ground state. This phenomenological explanation is in full accordance with our numerical many-body calculations described above. Obviously, it cannot predict the exact value of the critical mass ratio $\mu_c$ which is surprisingly small. Let us also note that this argumentation is also in full agreement with the mechanism of the spatial separation induced by repulsions in mass-imbalanced systems described in \cite{Pecak2016Separation}.

To make the analysis as complete as possible, we also discuss the inter-component correlations in terms of the von Neumann entropy which has been successfully used for the bosonic system also in the context of the mass imbalance~\cite{garcia2016entanglement}.
In contrast to the noise correlation which is based on the two-particle reduced density matrix, the von Neumann entropy is calculated from the reduced density matrix of the whole component. Therefore, it quantifies the total amount of correlations between components. It is defined straightforwardly as
\begin{equation}
{\cal S} = -\mathrm{Tr}\left(\hat\rho_\uparrow\ln\hat\rho_\uparrow\right),
\end{equation}
where $\hat\rho_\uparrow=\mathrm{Tr}_\downarrow\left(|G_0\rangle\langle G_0|\right)$ is the reduced density matrix of a chosen component calculated by tracing out the remaining component's degrees of freedom. As it is shown in the bottom row of Fig.~\ref{fig4}, the behavior of the von Neumann entropy is in full agreement with predictions based on the noise correlation. With increasing mass ratio $\mu$ the entropy rapidly decreases and after crossing some $\mu_c$ it slowly saturates on a small non-zero value. When comparing the von Neumann entropy $\cal S$ to the Frobenius distance $||{\cal G}_\tau||$ we see that these both quantities behave in a very similar way. It may suggest that all inter-component correlations are encoded mostly in the two-particle ones which probably dominate in the attractively interacting system. 

\section{Role of the Quantum Statistics}\label{sec:boseFermi}
\begin{figure}[t]
\centering
\includegraphics[width=\columnwidth]{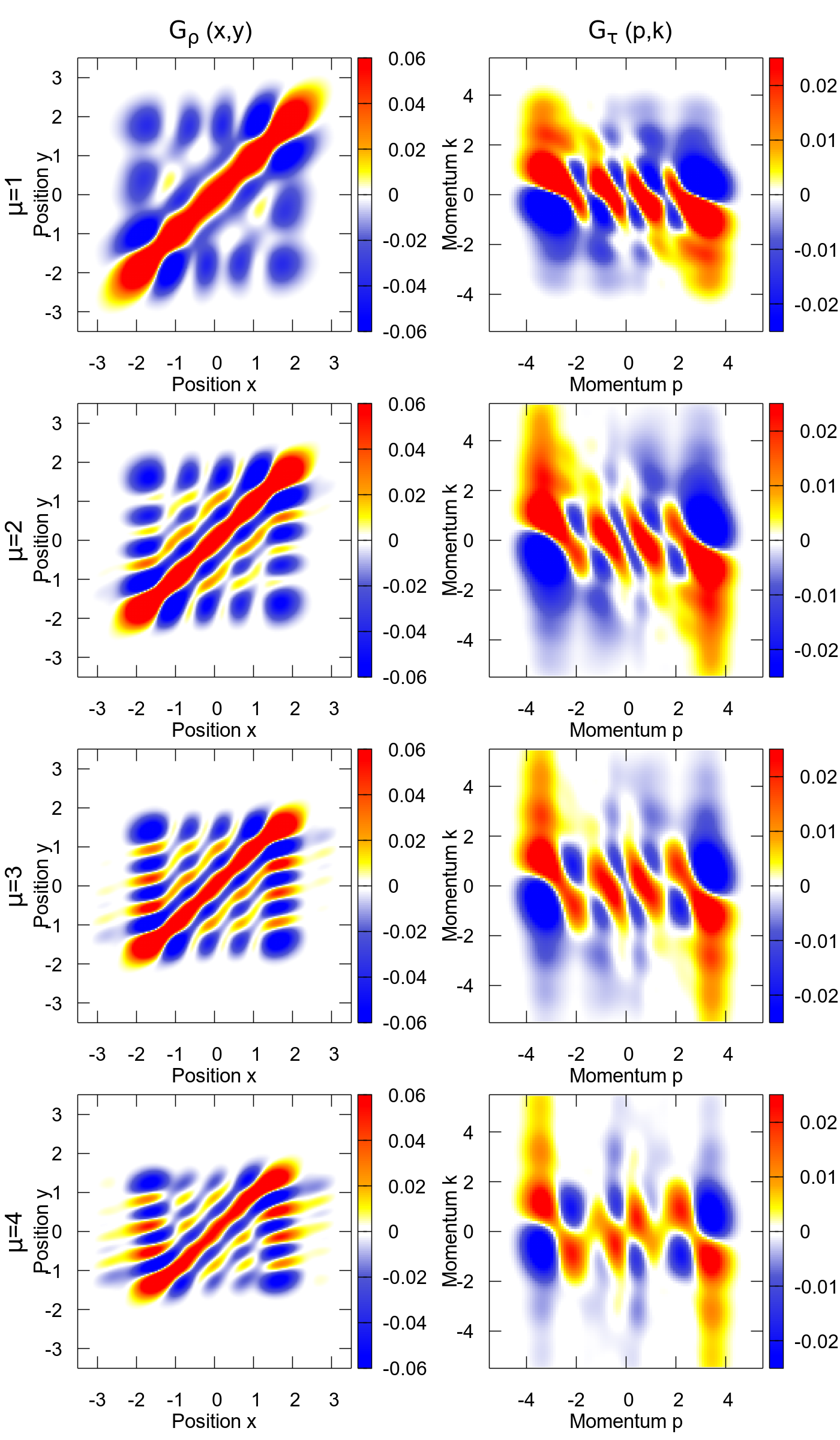}
\caption{
The noise correlation ${\cal G}_\rho$ and ${\cal G}_\tau$ calculated for the strongly attractive Bose-Fermi system ($g=-5$) described by the Hamiltonian \eqref{eq:hamBF} in the regime of strong repulsions between bosons {($g'=8$)} for $N_B=4$ bosons and $N_F=4$ fermions. While the inter-component correlations in the position domain are very similar to those obtained for fermionic mixtures, the anti-correlations in the momentum domain predicted previously are significantly destroyed even in the equal mass case ($\mu=1$). Position and momentum are measured in natural units of the harmonic oscillator, $\sqrt{\hbar/m_\downarrow\omega}$ and $\sqrt{\hbar m_\downarrow\omega}$, respectively.
}
\label{fig5}
\end{figure}
Finally, let us also discuss the role of the quantum statistics in forming inter-component correlations in the system studied. This analysis can be done systematically by changing one of the fermionic components to the bosonic one with the same number of particles and masses. In such a case the system is described by the modified Hamiltonian of the form:
\begin{multline} \label{eq:hamBF}
\hat{\cal H} = \int\!\!\mathrm{d}x\,\left[\hat{\Psi}^\dag(x)H_F\hat{\Psi}(x)+\hat{\Phi}^\dag(x)H_B\hat{\Phi}(x)\right]  \\
 + g\int\!\! \mathrm{d}x\,\hat{\Psi}^\dag(x)\hat{\Phi}^\dag(x)\hat{\Phi}(x)\hat{\Psi}(x) \\
 + g'\int\!\! \mathrm{d}x\,\hat{\Phi}^\dag(x)\hat{\Phi}^\dag(x)\hat{\Phi}(x)\hat{\Phi}(x).
\end{multline}
Here, the single-particle Hamiltonians $H_F$ and $H_B$ are given by equations \eqref{1PHam}, respectively (we replace spin-$\uparrow$ fermions described by the field $\hat\Psi_\uparrow(x)$ by interacting spinless bosons described by the field $\hat\Phi(x)$). The bosonic field operator $\hat{\Phi}(x)$ obeys the standard commutation relations $[\hat\Phi(x),\hat\Phi^\dagger(x')]=\delta(x-x')$ and $[\hat\Phi(x),\hat\Phi(x')]=0$.

The additional term in the Hamiltonian \eqref{eq:hamBF} which is proportional to $g'$ describes mutual interactions between bosons. To mimic the fermionic nature of bosons, in the following one assumes that $g'$ tends to infinite repulsions, {\it i.e.}, according to the Bose-Fermi mapping there exists a one-to-one correspondence between bosonic and fermionic wave functions in the component with replaced statistics~\cite{Girardeau1960}. In consequence, the spatial densities of infinitely repelling bosons and noninteracting fermions are exactly the same. Note, however, that there is a significant difference when density distributions of momenta are compared. This theoretical prediction was recently observed experimentally for two distinguishable fermions \cite{zurn2012fermionization} as well as for bosons confined in elongated traps~\cite{paredes2004tonks,kinoshita2004observation}. Of course, it is not possible to set $g'\rightarrow\infty$ in the numerical approach used. However, we checked that in the case of four bosons setting $g'=8$ appropriately mimics very strong repulsions for bosons and we use this value as the benchmark of infinite repulsions (see Appendix~\ref{appendixBosons} for details).

In the limit of infinite repulsions between bosons ($g'\rightarrow\infty$), the only difference between the two systems studied (modeled by Hamiltonians \eqref{eq:ham} and \eqref{eq:hamBF}) lies in the symmetry of the many-body wave function under exchange of two heavy particles. Despite this fact, the inter-component correlations (in the momentum domain) forced by attractive forces have significantly different properties. As it is seen in Fig.~\ref{fig5}, the noise correlations in the position domain for Bose-Fermi and Fermi-Fermi mixtures are very similar independently of the mass ratio $\mu$. This observation is a direct manifestation of the mapping mentioned above. However, in the momentum domain, the correlations described by the noise ${\cal G}_\tau$ are completely different. Even for the equal mass case $\mu=1$ the evident anti-correlation of momenta, previously clearly visible for fermions, is smeared and destroyed. 
This observation strongly suggests that the fermionic statistics present simultaneously in both components is crucial in building strong pairing (anti-correlations in momenta) in the system.

\section{Conclusion}\label{sec:conclusion}
To conclude, in this paper we discussed the properties of a two-component mixture of a few ultra-cold atoms in a one-dimensional harmonic trap with attractive mutual interactions. We focus on the inter-component correlations in terms of the noise correlation which effectively filters out single-particle features of the system from the two-body densities. In this way, we show that inter-component correlations strongly depend on the mass ratio between the atoms forming individual components. When the mass ratio is above $\mu_c\approx2$, the many-body ground state of the system undergoes a specific transition of its structure and it can be viewed as an almost perfect product of the noninteracting ground state of the heavier component and some well-defined state of the lighter particles. In consequence, inter-component correlations are strongly suppressed and are almost insensitive to the strength of attractive mutual interactions. 
Our numerical calculations predict a surprisingly small value of the critical mass (much below the ratio for a K-Li mixture, $\mu=40/6$, and Dy-K mixture, $\mu=161/40$) at which the transition occurs.
It is much smaller than for an impurity problem in the bosonic system studied recently~\cite{garcia2016entanglement}.

In addition, by studying two-component Bose-Fermi mixtures, we show that the quantum statistics play a crucial role in forming inter-component correlations. In this kind of a system, the anti-correlation of particles is strongly disturbed even in the system of equal mass components.

Since the noise correlation can be measured in nowadays experiments~\cite{bergschneider2018correlations}, our results may shed some light on incoming experiments with attractively interacting few-body systems.
Our analysis is quite general and it might be also important for building our understanding of different condensed matter problems related to the 'few' to 'many' crossover or unconventional superconductivity which originates in a pairing of different-mass fermions~\cite{HFsystems1984,matsuda2007fulde,kinnunen2018fulde}. For the same-mass fermions, the Fermi surfaces of both components match each other.
Whenever the Fermi surfaces do not match perfectly, a non-zero net momentum of pairs together with unconventional correlations may appear in the system.
A similar mechanism can occur for equal-mass systems with different trapping frequencies.

\section*{Acknowledgements}
We would like to thank Remigiusz Augusiak for valuable comments and Jacek Dobrzyniecki for a thorough reading of the manuscript.
This work was supported by the (Polish) National Science Center Grants No. 2016/21/N/ST2/03315 (DP) and 2016/22/E/ST2/00555 (TS). Numerical calculations were partially carried out in the 
Interdisciplinary Centre for Mathematical and Computational Modelling, University of Warsaw (ICM) under the computational grant No. G75-6.

\appendix
\section{Numerical convergence}~\label{AppendixConv}
\begin{figure}
\centering
\includegraphics[width=\columnwidth]{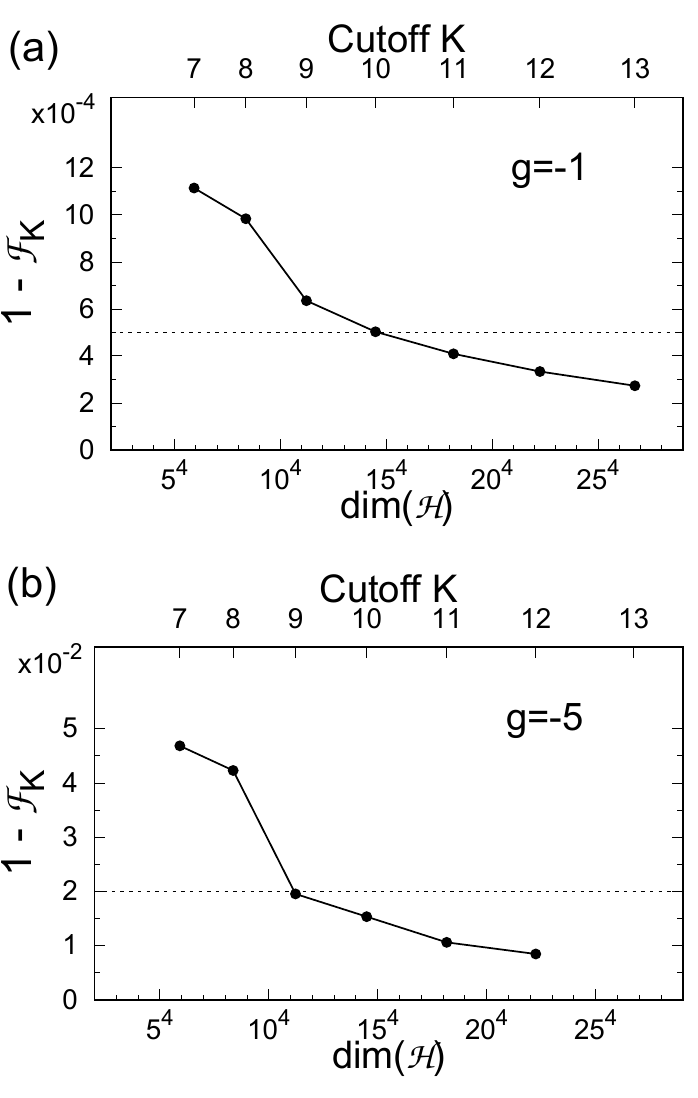}
\caption{
The successive fidelity ${\cal F}_K$ of the many-body ground state of $N_\uparrow=N_\downarrow=4$ equal-mass fermions obtained by the exact diagonalization for different values of the cut-off $K$ and two interaction strength (a) $g=-1$ and (b) $g=-5$. For the cut-off large enough the fidelity stabilizes (horizontal dashed line). Note that for better visibility, we use a nonlinear scaling on the horizontal axis.
}
\label{Fig6}
\end{figure}

\begin{figure}
\centering
\includegraphics[width=\columnwidth]{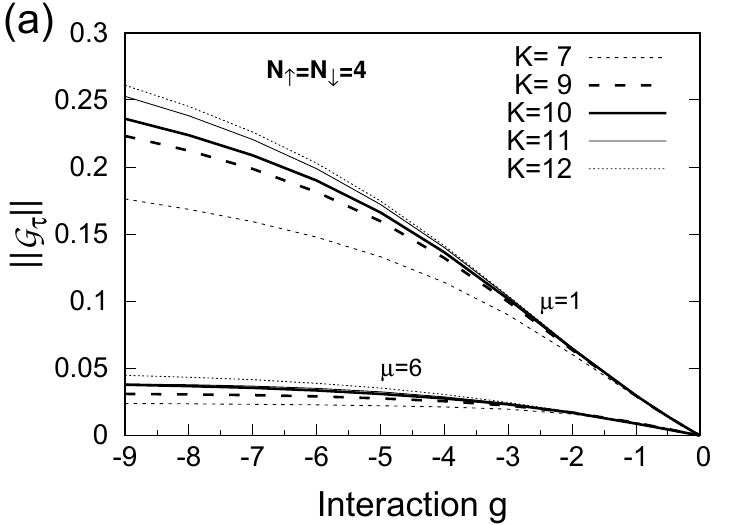}
\includegraphics[width=\columnwidth]{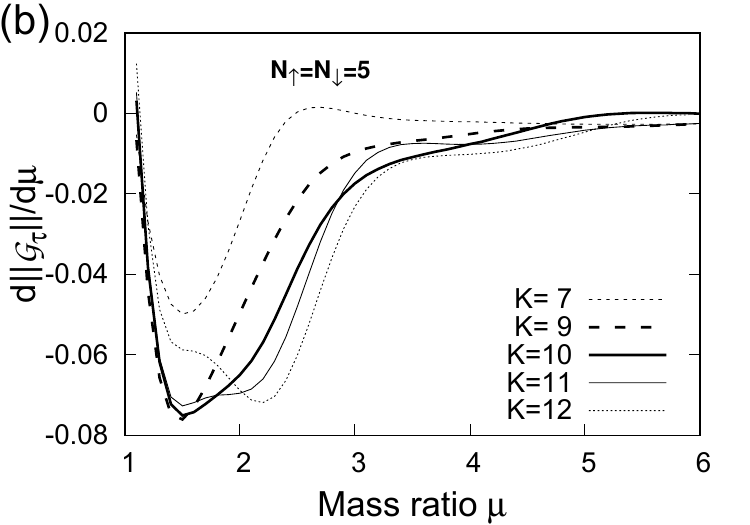}
\caption{Dependence of different quantities on the cut-off $K$. (a) The Frobenius distance $||{\cal G}_\tau||$ calculated for $N_\uparrow=N_\downarrow=4$ and two different mass ratios $\mu=1$ and $\mu=6$ (compare with Fig.~\ref{fig3}a). (b) A derivative of the Frobenius distance $\mathrm{d}||{\cal G}_\tau||/\mathrm{d}\mu$ calculated for $N_\uparrow=N_\downarrow=5$ and $g=-5$ (compare with the right middle plot in Fig.~\ref{fig4}).}
\label{Fig8}
\end{figure}

The convergence of the numerical method is ascertained by checking the successive fidelity of the many-body ground-state \cite{Pecak2017com}. The many-body Fock basis $\{|F_j\rangle\}$ is built from the lowest $K$ single-particle orbitals $\phi_{i\sigma}(x)$, $i\in\{0,\ldots,K-1\}$.  After numerical diagonalization of the many-body Hamiltonian \eqref{eq:hamSQ} one obtains the many-body ground state $|G^{\{K\}}_0\rangle$ and its energy $E^{\{K\}}_0$. By performing calculations for successive values of the cut-off $K$ we calculate the successive fidelity defined as
\begin{equation}
{\cal F}_K = |\langle G^{\{K-1\}}_0|G^{\{K\}}_0\rangle|.
\end{equation}
In Fig.~\ref{Fig6} we plot the difference $1-{\cal F}_K$ as a function of the cut-off $K$ for the system of $N_\uparrow=N_\downarrow=4$ fermions ($\mu=1$) and two different interaction strengths $g=-1$ and $g=-5$ (Fig.~\ref{Fig6}a and Fig.~\ref{Fig6}b, respectively). For convenience, we also display the corresponding sizes of the cropped many-body Hilbert space. We assume that the ground-state is found with sufficient accuracy if the changes of the fidelity are stabilized with the increasing cut-off (horizontal lines in Fig.~\ref{Fig6}). Note, that we use a nonlinear scaling on the horizontal axis, {\it i.e.}, the fidelity changes very slowly (in the stabilization region) with an increasing dimension of the Hilbert space. In these cases no significant changes of single- and two-particle densities with increasing cut-off are visible. The differences are also not significant when other quantities discussed are considered. In Fig.~\ref{Fig8}a we show the Frobenius distance $||{\cal G}_\tau||$ for $\mu=1$ and $\mu=6$ calculated with different cut-off $K$. As it is seen, in the range of attractions considered ($|g|\lessapprox 5$) the final result is well converged for $K\ge 10$. The situation is less obvious in the case of a derivative of the Frobenius distance $\mathrm{d}||{\cal G}_\tau||/\mathrm{d}\mu$ (middle row in Fig.~\ref{fig4}) which is much more sensitive to any changes of the Fock basis (see example for $N_\uparrow=N_\downarrow=5$ and $g=-5$ case in Fig.~\ref{Fig8}b). However, all the curves with $K\ge 10$  unambiguously support the observation that the ground state of the system undergoes a specific transition at around $\mu_c\approx 2$. 

\section{Fermionization limit}~\label{appendixBosons}
\begin{figure}
\centering
\includegraphics[width=\columnwidth]{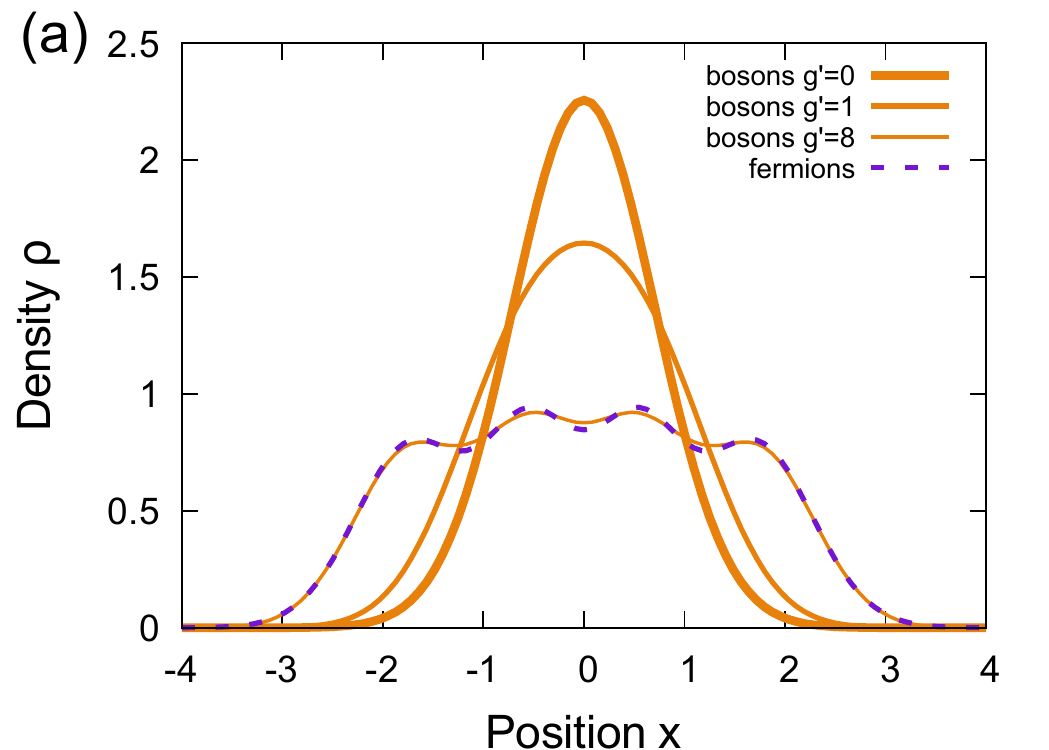}
\includegraphics[width=\columnwidth]{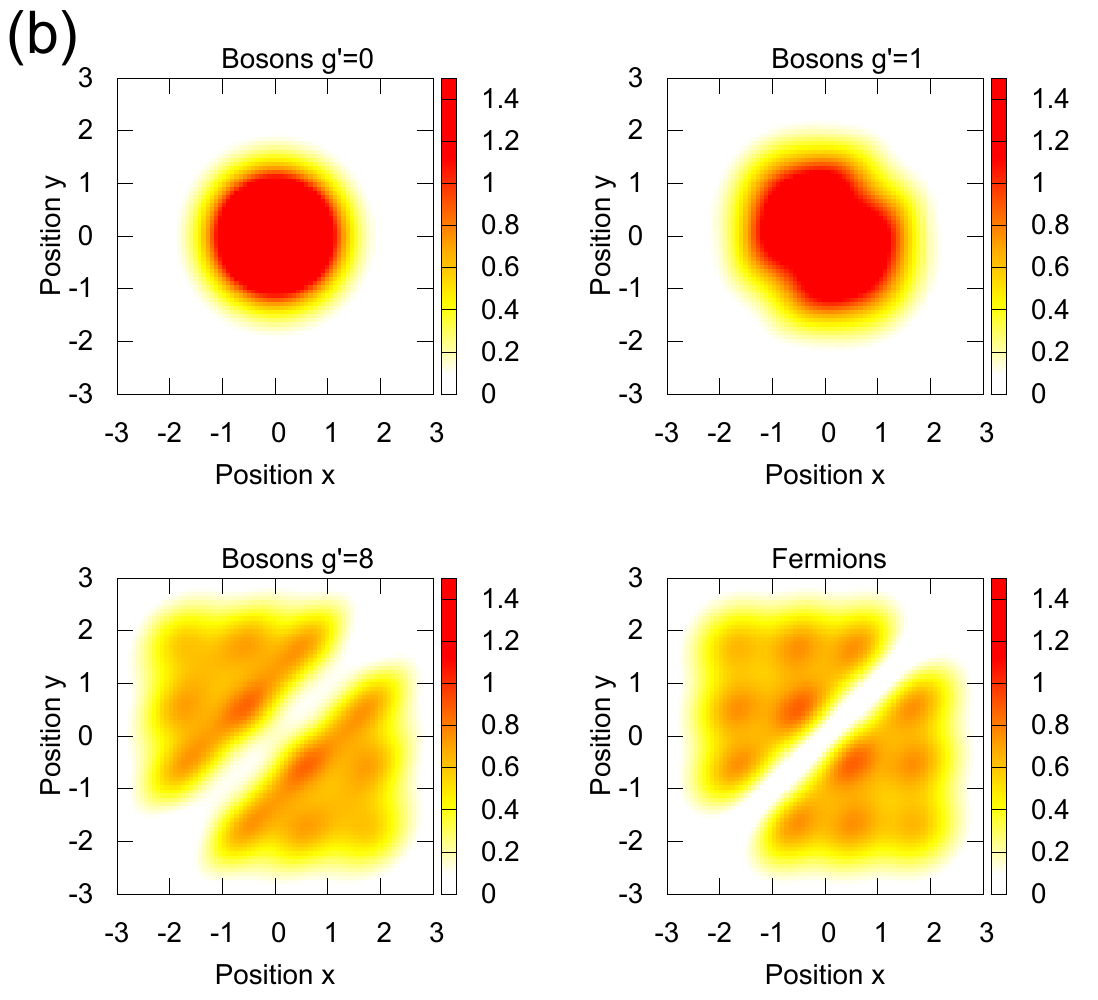}
\caption{
Comparison of the spatial properties of $N_F=4$ noninteracting fermions and the same number of repelling bosons for different interaction strengths $g'$. (a) The single-particle density distributions \eqref{rhoBF} for noninteracting fermions and bosons with different repulsions. (b) The two-particle density distributions \eqref{rho2BF} of corresponding four-particle systems. Note that in the case of $g'=8$ the single- and two-particle distributions are very close to the corresponding distributions of noninteracting fermions.
}
\label{Fig7}
\end{figure}
In the case of mixed statistics mixtures (bosons and fermions) discussed in Sec.~\ref{sec:boseFermi}, we assumed that the infinite repulsion limit between bosons ($g'\rightarrow\infty$) is appropriately mimicked by the finite value of interactions $g'=8$. To clarify this assumption we compare the ground-state spatial properties of $N_F=4$ noninteracting fermions (having the same spatial properties as infinitely repelling bosons) to $N_B=4$ bosons with different mutual interactions, $g'\in\{0, 1, 8\}$. Assuming that the ground state of $N_B$ interacting bosons ($N_F$ fermions) is $|G_B\rangle$ ($|G_F\rangle$) one defines the single-particle density as
\begin{subequations} \label{rhoBF}
\begin{align}
\rho^{(1)}_B(x)&=\langle G_B|\hat\Phi^\dagger(x)\hat\Phi(x)|G_B\rangle, \\
\rho^{(1)}_F(x) &= \langle G_F|\hat\Psi^\dagger(x)\hat\Psi(x)|G_F\rangle
\end{align}
\end{subequations}
for bosons and fermions, respectively. These distributions are shown in Fig.~\ref{Fig7}a. Similarly, the two-particle density distributions for the same systems of bosons and fermions are shown in Fig.~\ref{Fig7}b and are defined as
\begin{subequations} \label{rho2BF}
\begin{align}
\rho^{(2)}_B(x;y)&=\langle G_B|\hat\Phi^\dagger(x)\Phi^\dagger(y)\hat\Phi(y)\hat\Phi(x)|G_B\rangle, \\
\rho^{(2)}_F(x;y) &= \langle G_F|\hat\Psi^\dagger(x)\Psi^\dagger(y)\hat\Psi(y)\hat\Psi(x)|G_F\rangle,
\end{align}
\end{subequations}
respectively. It is clearly seen that the distributions of noninteracting fermions and noninteracting or weakly interacting bosons are essentially different. However, for strong repulsions the bosonic system undergoes fermionization and for $g'=8$ its spatial distributions become very close to corresponding distributions of the noninteracting fermionic system. This observation supports our assumption that the system of $N_B=4$ bosons interacting with the strength $g'=8$ can be safely treated as a benchmark of infinite repulsion.

\bibliographystyle{apsrev4-1}
\bibliography{bibtex}

\begin{thebibliography}{71}%
\makeatletter
\providecommand \@ifxundefined [1]{%
 \@ifx{#1\undefined}
}%
\providecommand \@ifnum [1]{%
 \ifnum #1\expandafter \@firstoftwo
 \else \expandafter \@secondoftwo
 \fi
}%
\providecommand \@ifx [1]{%
 \ifx #1\expandafter \@firstoftwo
 \else \expandafter \@secondoftwo
 \fi
}%
\providecommand \natexlab [1]{#1}%
\providecommand \enquote  [1]{``#1''}%
\providecommand \bibnamefont  [1]{#1}%
\providecommand \bibfnamefont [1]{#1}%
\providecommand \citenamefont [1]{#1}%
\providecommand \href@noop [0]{\@secondoftwo}%
\providecommand \href [0]{\begingroup \@sanitize@url \@href}%
\providecommand \@href[1]{\@@startlink{#1}\@@href}%
\providecommand \@@href[1]{\endgroup#1\@@endlink}%
\providecommand \@sanitize@url [0]{\catcode `\\12\catcode `\$12\catcode
  `\&12\catcode `\#12\catcode `\^12\catcode `\_12\catcode `\%12\relax}%
\providecommand \@@startlink[1]{}%
\providecommand \@@endlink[0]{}%
\providecommand \url  [0]{\begingroup\@sanitize@url \@url }%
\providecommand \@url [1]{\endgroup\@href {#1}{\urlprefix }}%
\providecommand \urlprefix  [0]{URL }%
\providecommand \Eprint [0]{\href }%
\providecommand \doibase [0]{http://dx.doi.org/}%
\providecommand \selectlanguage [0]{\@gobble}%
\providecommand \bibinfo  [0]{\@secondoftwo}%
\providecommand \bibfield  [0]{\@secondoftwo}%
\providecommand \translation [1]{[#1]}%
\providecommand \BibitemOpen [0]{}%
\providecommand \bibitemStop [0]{}%
\providecommand \bibitemNoStop [0]{.\EOS\space}%
\providecommand \EOS [0]{\spacefactor3000\relax}%
\providecommand \BibitemShut  [1]{\csname bibitem#1\endcsname}%
\let\auto@bib@innerbib\@empty
\bibitem [{\citenamefont {Wenz}\ \emph {et~al.}(2013)\citenamefont {Wenz},
  \citenamefont {Z{\"u}rn}, \citenamefont {Murmann}, \citenamefont {Brouzos},
  \citenamefont {Lompe},\ and\ \citenamefont {Jochim}}]{wenz2013fewToMany}%
  \BibitemOpen
  \bibfield  {author} {\bibinfo {author} {\bibfnamefont {A.}~\bibnamefont
  {Wenz}}, \bibinfo {author} {\bibfnamefont {G.}~\bibnamefont {Z{\"u}rn}},
  \bibinfo {author} {\bibfnamefont {S.}~\bibnamefont {Murmann}}, \bibinfo
  {author} {\bibfnamefont {I.}~\bibnamefont {Brouzos}}, \bibinfo {author}
  {\bibfnamefont {T.}~\bibnamefont {Lompe}}, \ and\ \bibinfo {author}
  {\bibfnamefont {S.}~\bibnamefont {Jochim}},\ }\href@noop {} {\bibfield
  {journal} {\bibinfo  {journal} {Science}\ }\textbf {\bibinfo {volume}
  {342}},\ \bibinfo {pages} {457} (\bibinfo {year} {2013})}\BibitemShut
  {NoStop}%
\bibitem [{\citenamefont {Z\"urn}\ \emph {et~al.}(2012)\citenamefont {Z\"urn},
  \citenamefont {Serwane}, \citenamefont {Lompe}, \citenamefont {Wenz},
  \citenamefont {Ries}, \citenamefont {Bohn},\ and\ \citenamefont
  {Jochim}}]{zurn2012fermionization}%
  \BibitemOpen
  \bibfield  {author} {\bibinfo {author} {\bibfnamefont {G.}~\bibnamefont
  {Z\"urn}}, \bibinfo {author} {\bibfnamefont {F.}~\bibnamefont {Serwane}},
  \bibinfo {author} {\bibfnamefont {T.}~\bibnamefont {Lompe}}, \bibinfo
  {author} {\bibfnamefont {A.~N.}\ \bibnamefont {Wenz}}, \bibinfo {author}
  {\bibfnamefont {M.~G.}\ \bibnamefont {Ries}}, \bibinfo {author}
  {\bibfnamefont {J.~E.}\ \bibnamefont {Bohn}}, \ and\ \bibinfo {author}
  {\bibfnamefont {S.}~\bibnamefont {Jochim}},\ }\href@noop {} {\bibfield
  {journal} {\bibinfo  {journal} {Phys. Rev. Lett.}\ }\textbf {\bibinfo
  {volume} {108}},\ \bibinfo {pages} {075303} (\bibinfo {year}
  {2012})}\BibitemShut {NoStop}%
\bibitem [{\citenamefont {Serwane}\ \emph {et~al.}(2011)\citenamefont
  {Serwane}, \citenamefont {Z{\"u}rn}, \citenamefont {Lompe}, \citenamefont
  {Ottenstein}, \citenamefont {Wenz},\ and\ \citenamefont
  {Jochim}}]{serwane2011deterministic}%
  \BibitemOpen
  \bibfield  {author} {\bibinfo {author} {\bibfnamefont {F.}~\bibnamefont
  {Serwane}}, \bibinfo {author} {\bibfnamefont {G.}~\bibnamefont {Z{\"u}rn}},
  \bibinfo {author} {\bibfnamefont {T.}~\bibnamefont {Lompe}}, \bibinfo
  {author} {\bibfnamefont {T.}~\bibnamefont {Ottenstein}}, \bibinfo {author}
  {\bibfnamefont {A.}~\bibnamefont {Wenz}}, \ and\ \bibinfo {author}
  {\bibfnamefont {S.}~\bibnamefont {Jochim}},\ }\href@noop {} {\bibfield
  {journal} {\bibinfo  {journal} {Science}\ }\textbf {\bibinfo {volume}
  {332}},\ \bibinfo {pages} {336} (\bibinfo {year} {2011})}\BibitemShut
  {NoStop}%
\bibitem [{\citenamefont {Murmann}\ \emph {et~al.}(2015)\citenamefont
  {Murmann}, \citenamefont {Deuretzbacher}, \citenamefont {Z\"{u}rn},
  \citenamefont {Bjerlin}, \citenamefont {Reimann}, \citenamefont {Santos},
  \citenamefont {Lompe},\ and\ \citenamefont
  {Jochim}}]{Murmann2015AntiferroSpinChain}%
  \BibitemOpen
  \bibfield  {author} {\bibinfo {author} {\bibfnamefont {S.}~\bibnamefont
  {Murmann}}, \bibinfo {author} {\bibfnamefont {F.}~\bibnamefont
  {Deuretzbacher}}, \bibinfo {author} {\bibfnamefont {G.}~\bibnamefont
  {Z\"{u}rn}}, \bibinfo {author} {\bibfnamefont {J.}~\bibnamefont {Bjerlin}},
  \bibinfo {author} {\bibfnamefont {S.~M.}\ \bibnamefont {Reimann}}, \bibinfo
  {author} {\bibfnamefont {L.}~\bibnamefont {Santos}}, \bibinfo {author}
  {\bibfnamefont {T.}~\bibnamefont {Lompe}}, \ and\ \bibinfo {author}
  {\bibfnamefont {S.}~\bibnamefont {Jochim}},\ }\href@noop {} {\bibfield
  {journal} {\bibinfo  {journal} {Phys. Rev. Lett.}\ }\textbf {\bibinfo
  {volume} {115}},\ \bibinfo {pages} {215301} (\bibinfo {year}
  {2015})}\BibitemShut {NoStop}%
\bibitem [{\citenamefont {Kaufman}\ \emph {et~al.}(2015)\citenamefont
  {Kaufman}, \citenamefont {Lester}, \citenamefont {Foss-Feig}, \citenamefont
  {Wall}, \citenamefont {Rey},\ and\ \citenamefont
  {Regal}}]{Kaufman2015Entangling}%
  \BibitemOpen
  \bibfield  {author} {\bibinfo {author} {\bibfnamefont {A.~M.}\ \bibnamefont
  {Kaufman}}, \bibinfo {author} {\bibfnamefont {B.~J.}\ \bibnamefont {Lester}},
  \bibinfo {author} {\bibfnamefont {M.}~\bibnamefont {Foss-Feig}}, \bibinfo
  {author} {\bibfnamefont {M.~L.}\ \bibnamefont {Wall}}, \bibinfo {author}
  {\bibfnamefont {A.~M.}\ \bibnamefont {Rey}}, \ and\ \bibinfo {author}
  {\bibfnamefont {C.~A.}\ \bibnamefont {Regal}},\ }\href@noop {} {\bibfield
  {journal} {\bibinfo  {journal} {Nature}\ }\textbf {\bibinfo {volume} {527}},\
  \bibinfo {pages} {208} (\bibinfo {year} {2015})}\BibitemShut {NoStop}%
\bibitem [{\citenamefont {Bergschneider}\ \emph
  {et~al.}(2018{\natexlab{a}})\citenamefont {Bergschneider}, \citenamefont
  {Klinkhamer}, \citenamefont {Becher}, \citenamefont {Klemt}, \citenamefont
  {Z\"urn}, \citenamefont {Preiss},\ and\ \citenamefont
  {Jochim}}]{Andrea2018Imaging}%
  \BibitemOpen
  \bibfield  {author} {\bibinfo {author} {\bibfnamefont {A.}~\bibnamefont
  {Bergschneider}}, \bibinfo {author} {\bibfnamefont {V.~M.}\ \bibnamefont
  {Klinkhamer}}, \bibinfo {author} {\bibfnamefont {J.~H.}\ \bibnamefont
  {Becher}}, \bibinfo {author} {\bibfnamefont {R.}~\bibnamefont {Klemt}},
  \bibinfo {author} {\bibfnamefont {G.}~\bibnamefont {Z\"urn}}, \bibinfo
  {author} {\bibfnamefont {P.~M.}\ \bibnamefont {Preiss}}, \ and\ \bibinfo
  {author} {\bibfnamefont {S.}~\bibnamefont {Jochim}},\ }\href@noop {}
  {\bibfield  {journal} {\bibinfo  {journal} {Phys. Rev. A}\ }\textbf {\bibinfo
  {volume} {97}},\ \bibinfo {pages} {063613} (\bibinfo {year}
  {2018}{\natexlab{a}})}\BibitemShut {NoStop}%
\bibitem [{\citenamefont {Z\"urn}\ \emph {et~al.}(2013)\citenamefont {Z\"urn},
  \citenamefont {Wenz}, \citenamefont {Murmann}, \citenamefont {Bergschneider},
  \citenamefont {Lompe},\ and\ \citenamefont {Jochim}}]{zurn2013Pairing}%
  \BibitemOpen
  \bibfield  {author} {\bibinfo {author} {\bibfnamefont {G.}~\bibnamefont
  {Z\"urn}}, \bibinfo {author} {\bibfnamefont {A.~N.}\ \bibnamefont {Wenz}},
  \bibinfo {author} {\bibfnamefont {S.}~\bibnamefont {Murmann}}, \bibinfo
  {author} {\bibfnamefont {A.}~\bibnamefont {Bergschneider}}, \bibinfo {author}
  {\bibfnamefont {T.}~\bibnamefont {Lompe}}, \ and\ \bibinfo {author}
  {\bibfnamefont {S.}~\bibnamefont {Jochim}},\ }\href@noop {} {\bibfield
  {journal} {\bibinfo  {journal} {Phys. Rev. Lett.}\ }\textbf {\bibinfo
  {volume} {111}},\ \bibinfo {pages} {175302} (\bibinfo {year}
  {2013})}\BibitemShut {NoStop}%
\bibitem [{\citenamefont {Bergschneider}\ \emph
  {et~al.}(2018{\natexlab{b}})\citenamefont {Bergschneider}, \citenamefont
  {Klinkhamer}, \citenamefont {Becher}, \citenamefont {Klemt}, \citenamefont
  {Palm}, \citenamefont {Z{\"u}rn}, \citenamefont {Jochim},\ and\ \citenamefont
  {Preiss}}]{bergschneider2018correlations}%
  \BibitemOpen
  \bibfield  {author} {\bibinfo {author} {\bibfnamefont {A.}~\bibnamefont
  {Bergschneider}}, \bibinfo {author} {\bibfnamefont {V.~M.}\ \bibnamefont
  {Klinkhamer}}, \bibinfo {author} {\bibfnamefont {J.~H.}\ \bibnamefont
  {Becher}}, \bibinfo {author} {\bibfnamefont {R.}~\bibnamefont {Klemt}},
  \bibinfo {author} {\bibfnamefont {L.}~\bibnamefont {Palm}}, \bibinfo {author}
  {\bibfnamefont {G.}~\bibnamefont {Z{\"u}rn}}, \bibinfo {author}
  {\bibfnamefont {S.}~\bibnamefont {Jochim}}, \ and\ \bibinfo {author}
  {\bibfnamefont {P.~M.}\ \bibnamefont {Preiss}},\ }\href@noop {} {\bibfield
  {journal} {\bibinfo  {journal} {arXiv preprint arXiv:1807.06405}\ } (\bibinfo
  {year} {2018}{\natexlab{b}})}\BibitemShut {NoStop}%
\bibitem [{\citenamefont {Bakr}\ \emph {et~al.}(2010)\citenamefont {Bakr},
  \citenamefont {Peng}, \citenamefont {Tai}, \citenamefont {Ma}, \citenamefont
  {Simon}, \citenamefont {Gillen}, \citenamefont {Foelling}, \citenamefont
  {Pollet},\ and\ \citenamefont {Greiner}}]{bakr2010probing}%
  \BibitemOpen
  \bibfield  {author} {\bibinfo {author} {\bibfnamefont {W.~S.}\ \bibnamefont
  {Bakr}}, \bibinfo {author} {\bibfnamefont {A.}~\bibnamefont {Peng}}, \bibinfo
  {author} {\bibfnamefont {M.~E.}\ \bibnamefont {Tai}}, \bibinfo {author}
  {\bibfnamefont {R.}~\bibnamefont {Ma}}, \bibinfo {author} {\bibfnamefont
  {J.}~\bibnamefont {Simon}}, \bibinfo {author} {\bibfnamefont {J.~I.}\
  \bibnamefont {Gillen}}, \bibinfo {author} {\bibfnamefont {S.}~\bibnamefont
  {Foelling}}, \bibinfo {author} {\bibfnamefont {L.}~\bibnamefont {Pollet}}, \
  and\ \bibinfo {author} {\bibfnamefont {M.}~\bibnamefont {Greiner}},\
  }\href@noop {} {\bibfield  {journal} {\bibinfo  {journal} {Science}\ }\textbf
  {\bibinfo {volume} {329}},\ \bibinfo {pages} {547} (\bibinfo {year}
  {2010})}\BibitemShut {NoStop}%
\bibitem [{\citenamefont {Sherson}\ \emph {et~al.}(2010)\citenamefont
  {Sherson}, \citenamefont {Weitenberg}, \citenamefont {Endres}, \citenamefont
  {Cheneau}, \citenamefont {Bloch},\ and\ \citenamefont
  {Kuhr}}]{sherson2010single}%
  \BibitemOpen
  \bibfield  {author} {\bibinfo {author} {\bibfnamefont {J.~F.}\ \bibnamefont
  {Sherson}}, \bibinfo {author} {\bibfnamefont {C.}~\bibnamefont {Weitenberg}},
  \bibinfo {author} {\bibfnamefont {M.}~\bibnamefont {Endres}}, \bibinfo
  {author} {\bibfnamefont {M.}~\bibnamefont {Cheneau}}, \bibinfo {author}
  {\bibfnamefont {I.}~\bibnamefont {Bloch}}, \ and\ \bibinfo {author}
  {\bibfnamefont {S.}~\bibnamefont {Kuhr}},\ }\href@noop {} {\bibfield
  {journal} {\bibinfo  {journal} {Nature}\ }\textbf {\bibinfo {volume} {467}},\
  \bibinfo {pages} {68} (\bibinfo {year} {2010})}\BibitemShut {NoStop}%
\bibitem [{\citenamefont {Omran}\ \emph {et~al.}(2015)\citenamefont {Omran},
  \citenamefont {Boll}, \citenamefont {Hilker}, \citenamefont {Kleinlein},
  \citenamefont {Salomon}, \citenamefont {Bloch},\ and\ \citenamefont
  {Gross}}]{OmranBloch2015PauliBlocking}%
  \BibitemOpen
  \bibfield  {author} {\bibinfo {author} {\bibfnamefont {A.}~\bibnamefont
  {Omran}}, \bibinfo {author} {\bibfnamefont {M.}~\bibnamefont {Boll}},
  \bibinfo {author} {\bibfnamefont {T.~A.}\ \bibnamefont {Hilker}}, \bibinfo
  {author} {\bibfnamefont {K.}~\bibnamefont {Kleinlein}}, \bibinfo {author}
  {\bibfnamefont {G.}~\bibnamefont {Salomon}}, \bibinfo {author} {\bibfnamefont
  {I.}~\bibnamefont {Bloch}}, \ and\ \bibinfo {author} {\bibfnamefont
  {C.}~\bibnamefont {Gross}},\ }\href@noop {} {\bibfield  {journal} {\bibinfo
  {journal} {Phys. Rev. Lett.}\ }\textbf {\bibinfo {volume} {115}},\ \bibinfo
  {pages} {263001} (\bibinfo {year} {2015})}\BibitemShut {NoStop}%
\bibitem [{\citenamefont {Cheuk}\ \emph {et~al.}(2015)\citenamefont {Cheuk},
  \citenamefont {Nichols}, \citenamefont {Okan}, \citenamefont {Gersdorf},
  \citenamefont {Ramasesh}, \citenamefont {Bakr}, \citenamefont {Lompe},\ and\
  \citenamefont {Zwierlein}}]{Cheuk2015FermionicMicroscope}%
  \BibitemOpen
  \bibfield  {author} {\bibinfo {author} {\bibfnamefont {L.~W.}\ \bibnamefont
  {Cheuk}}, \bibinfo {author} {\bibfnamefont {M.~A.}\ \bibnamefont {Nichols}},
  \bibinfo {author} {\bibfnamefont {M.}~\bibnamefont {Okan}}, \bibinfo {author}
  {\bibfnamefont {T.}~\bibnamefont {Gersdorf}}, \bibinfo {author}
  {\bibfnamefont {V.~V.}\ \bibnamefont {Ramasesh}}, \bibinfo {author}
  {\bibfnamefont {W.~S.}\ \bibnamefont {Bakr}}, \bibinfo {author}
  {\bibfnamefont {T.}~\bibnamefont {Lompe}}, \ and\ \bibinfo {author}
  {\bibfnamefont {M.~W.}\ \bibnamefont {Zwierlein}},\ }\href@noop {} {\bibfield
   {journal} {\bibinfo  {journal} {Phys. Rev. Lett.}\ }\textbf {\bibinfo
  {volume} {114}},\ \bibinfo {pages} {193001} (\bibinfo {year}
  {2015})}\BibitemShut {NoStop}%
\bibitem [{\citenamefont {Cheuk}\ \emph
  {et~al.}(2016{\natexlab{a}})\citenamefont {Cheuk}, \citenamefont {Nichols},
  \citenamefont {Lawrence}, \citenamefont {Okan}, \citenamefont {Zhang},\ and\
  \citenamefont {Zwierlein}}]{Cheuk2016MottMicroscope}%
  \BibitemOpen
  \bibfield  {author} {\bibinfo {author} {\bibfnamefont {L.~W.}\ \bibnamefont
  {Cheuk}}, \bibinfo {author} {\bibfnamefont {M.~A.}\ \bibnamefont {Nichols}},
  \bibinfo {author} {\bibfnamefont {K.~R.}\ \bibnamefont {Lawrence}}, \bibinfo
  {author} {\bibfnamefont {M.}~\bibnamefont {Okan}}, \bibinfo {author}
  {\bibfnamefont {H.}~\bibnamefont {Zhang}}, \ and\ \bibinfo {author}
  {\bibfnamefont {M.~W.}\ \bibnamefont {Zwierlein}},\ }\href@noop {} {\bibfield
   {journal} {\bibinfo  {journal} {Phys. Rev. Lett.}\ }\textbf {\bibinfo
  {volume} {116}},\ \bibinfo {pages} {235301} (\bibinfo {year}
  {2016}{\natexlab{a}})}\BibitemShut {NoStop}%
\bibitem [{\citenamefont {Cheuk}\ \emph
  {et~al.}(2016{\natexlab{b}})\citenamefont {Cheuk}, \citenamefont {Nichols},
  \citenamefont {Lawrence}, \citenamefont {Okan}, \citenamefont {Zhang},
  \citenamefont {Khatami}, \citenamefont {Trivedi}, \citenamefont {Paiva},
  \citenamefont {Rigol},\ and\ \citenamefont
  {Zwierlein}}]{Cheuk2016Correlations}%
  \BibitemOpen
  \bibfield  {author} {\bibinfo {author} {\bibfnamefont {L.~W.}\ \bibnamefont
  {Cheuk}}, \bibinfo {author} {\bibfnamefont {M.~A.}\ \bibnamefont {Nichols}},
  \bibinfo {author} {\bibfnamefont {K.~R.}\ \bibnamefont {Lawrence}}, \bibinfo
  {author} {\bibfnamefont {M.}~\bibnamefont {Okan}}, \bibinfo {author}
  {\bibfnamefont {H.}~\bibnamefont {Zhang}}, \bibinfo {author} {\bibfnamefont
  {E.}~\bibnamefont {Khatami}}, \bibinfo {author} {\bibfnamefont
  {N.}~\bibnamefont {Trivedi}}, \bibinfo {author} {\bibfnamefont
  {T.}~\bibnamefont {Paiva}}, \bibinfo {author} {\bibfnamefont
  {M.}~\bibnamefont {Rigol}}, \ and\ \bibinfo {author} {\bibfnamefont {M.~W.}\
  \bibnamefont {Zwierlein}},\ }\href@noop {} {\bibfield  {journal} {\bibinfo
  {journal} {Science}\ }\textbf {\bibinfo {volume} {353}},\ \bibinfo {pages}
  {1260} (\bibinfo {year} {2016}{\natexlab{b}})}\BibitemShut {NoStop}%
\bibitem [{\citenamefont {Parsons}\ \emph {et~al.}(2015)\citenamefont
  {Parsons}, \citenamefont {Huber}, \citenamefont {Mazurenko}, \citenamefont
  {Chiu}, \citenamefont {Setiawan}, \citenamefont {Wooley-Brown}, \citenamefont
  {Blatt},\ and\ \citenamefont {Greiner}}]{ParsonsPRL2015}%
  \BibitemOpen
  \bibfield  {author} {\bibinfo {author} {\bibfnamefont {M.~F.}\ \bibnamefont
  {Parsons}}, \bibinfo {author} {\bibfnamefont {F.}~\bibnamefont {Huber}},
  \bibinfo {author} {\bibfnamefont {A.}~\bibnamefont {Mazurenko}}, \bibinfo
  {author} {\bibfnamefont {C.~S.}\ \bibnamefont {Chiu}}, \bibinfo {author}
  {\bibfnamefont {W.}~\bibnamefont {Setiawan}}, \bibinfo {author}
  {\bibfnamefont {K.}~\bibnamefont {Wooley-Brown}}, \bibinfo {author}
  {\bibfnamefont {S.}~\bibnamefont {Blatt}}, \ and\ \bibinfo {author}
  {\bibfnamefont {M.}~\bibnamefont {Greiner}},\ }\href@noop {} {\bibfield
  {journal} {\bibinfo  {journal} {Phys. Rev. Lett.}\ }\textbf {\bibinfo
  {volume} {114}},\ \bibinfo {pages} {213002} (\bibinfo {year}
  {2015})}\BibitemShut {NoStop}%
\bibitem [{\citenamefont {Edge}\ \emph {et~al.}(2015)\citenamefont {Edge},
  \citenamefont {Anderson}, \citenamefont {Jervis}, \citenamefont {McKay},
  \citenamefont {Day}, \citenamefont {Trotzky},\ and\ \citenamefont
  {Thywissen}}]{EdgePRA2015ImagingFermions}%
  \BibitemOpen
  \bibfield  {author} {\bibinfo {author} {\bibfnamefont {G.~J.~A.}\
  \bibnamefont {Edge}}, \bibinfo {author} {\bibfnamefont {R.}~\bibnamefont
  {Anderson}}, \bibinfo {author} {\bibfnamefont {D.}~\bibnamefont {Jervis}},
  \bibinfo {author} {\bibfnamefont {D.~C.}\ \bibnamefont {McKay}}, \bibinfo
  {author} {\bibfnamefont {R.}~\bibnamefont {Day}}, \bibinfo {author}
  {\bibfnamefont {S.}~\bibnamefont {Trotzky}}, \ and\ \bibinfo {author}
  {\bibfnamefont {J.~H.}\ \bibnamefont {Thywissen}},\ }\href@noop {} {\bibfield
   {journal} {\bibinfo  {journal} {Phys. Rev. A}\ }\textbf {\bibinfo {volume}
  {92}},\ \bibinfo {pages} {063406} (\bibinfo {year} {2015})}\BibitemShut
  {NoStop}%
\bibitem [{\citenamefont {Haller}\ \emph {et~al.}(2015)\citenamefont {Haller},
  \citenamefont {Hudson}, \citenamefont {Kelly}, \citenamefont {Cotta},
  \citenamefont {Peaudecerf}, \citenamefont {Bruce},\ and\ \citenamefont
  {Kuhr}}]{haller2015single}%
  \BibitemOpen
  \bibfield  {author} {\bibinfo {author} {\bibfnamefont {E.}~\bibnamefont
  {Haller}}, \bibinfo {author} {\bibfnamefont {J.}~\bibnamefont {Hudson}},
  \bibinfo {author} {\bibfnamefont {A.}~\bibnamefont {Kelly}}, \bibinfo
  {author} {\bibfnamefont {D.~A.}\ \bibnamefont {Cotta}}, \bibinfo {author}
  {\bibfnamefont {B.}~\bibnamefont {Peaudecerf}}, \bibinfo {author}
  {\bibfnamefont {G.~D.}\ \bibnamefont {Bruce}}, \ and\ \bibinfo {author}
  {\bibfnamefont {S.}~\bibnamefont {Kuhr}},\ }\href@noop {} {\bibfield
  {journal} {\bibinfo  {journal} {Nature Physics}\ }\textbf {\bibinfo {volume}
  {11}},\ \bibinfo {pages} {738} (\bibinfo {year} {2015})}\BibitemShut
  {NoStop}%
\bibitem [{\citenamefont {Brouzos}\ and\ \citenamefont
  {Schmelcher}(2013)}]{brouzos2013two}%
  \BibitemOpen
  \bibfield  {author} {\bibinfo {author} {\bibfnamefont {I.}~\bibnamefont
  {Brouzos}}\ and\ \bibinfo {author} {\bibfnamefont {P.}~\bibnamefont
  {Schmelcher}},\ }\href@noop {} {\bibfield  {journal} {\bibinfo  {journal}
  {Physical Review A}\ }\textbf {\bibinfo {volume} {87}},\ \bibinfo {pages}
  {023605} (\bibinfo {year} {2013})}\BibitemShut {NoStop}%
\bibitem [{\citenamefont {Sowi\'{n}ski}\ \emph {et~al.}(2013)\citenamefont
  {Sowi\'{n}ski}, \citenamefont {Grass}, \citenamefont {Dutta},\ and\
  \citenamefont {Lewenstein}}]{SowinskiGrass2013FewInteracting}%
  \BibitemOpen
  \bibfield  {author} {\bibinfo {author} {\bibfnamefont {T.}~\bibnamefont
  {Sowi\'{n}ski}}, \bibinfo {author} {\bibfnamefont {T.}~\bibnamefont {Grass}},
  \bibinfo {author} {\bibfnamefont {O.}~\bibnamefont {Dutta}}, \ and\ \bibinfo
  {author} {\bibfnamefont {M.}~\bibnamefont {Lewenstein}},\ }\href@noop {}
  {\bibfield  {journal} {\bibinfo  {journal} {Phys. Rev. A}\ }\textbf {\bibinfo
  {volume} {88}},\ \bibinfo {pages} {033607} (\bibinfo {year}
  {2013})}\BibitemShut {NoStop}%
\bibitem [{\citenamefont {Gharashi}\ and\ \citenamefont
  {Blume}(2013)}]{Gharashi2013UpperBranchCorrelations}%
  \BibitemOpen
  \bibfield  {author} {\bibinfo {author} {\bibfnamefont {S.~E.}\ \bibnamefont
  {Gharashi}}\ and\ \bibinfo {author} {\bibfnamefont {D.}~\bibnamefont
  {Blume}},\ }\href@noop {} {\bibfield  {journal} {\bibinfo  {journal} {Phys.
  Rev. Lett.}\ }\textbf {\bibinfo {volume} {111}},\ \bibinfo {pages} {045302}
  (\bibinfo {year} {2013})}\BibitemShut {NoStop}%
\bibitem [{\citenamefont {Garc\'{\i}a-March}\ \emph
  {et~al.}(2014{\natexlab{a}})\citenamefont {Garc\'{\i}a-March}, \citenamefont
  {Juli\'{a}-D\'{\i}az}, \citenamefont {Astrakharchik}, \citenamefont {Busch},
  \citenamefont {Boronat},\ and\ \citenamefont
  {Polls}}]{GarciaMarch2014Localization}%
  \BibitemOpen
  \bibfield  {author} {\bibinfo {author} {\bibfnamefont {M.~A.}\ \bibnamefont
  {Garc\'{\i}a-March}}, \bibinfo {author} {\bibfnamefont {B.}~\bibnamefont
  {Juli\'{a}-D\'{\i}az}}, \bibinfo {author} {\bibfnamefont {G.~E.}\
  \bibnamefont {Astrakharchik}}, \bibinfo {author} {\bibfnamefont
  {T.}~\bibnamefont {Busch}}, \bibinfo {author} {\bibfnamefont
  {J.}~\bibnamefont {Boronat}}, \ and\ \bibinfo {author} {\bibfnamefont
  {A.}~\bibnamefont {Polls}},\ }\href@noop {} {\bibfield  {journal} {\bibinfo
  {journal} {New J. Phys.}\ }\textbf {\bibinfo {volume} {16}},\ \bibinfo
  {pages} {103004} (\bibinfo {year} {2014}{\natexlab{a}})}\BibitemShut
  {NoStop}%
\bibitem [{\citenamefont {Fogarty}\ \emph {et~al.}(2018)\citenamefont
  {Fogarty}, \citenamefont {Ruks}, \citenamefont {Li},\ and\ \citenamefont
  {Busch}}]{fogarty2018fast}%
  \BibitemOpen
  \bibfield  {author} {\bibinfo {author} {\bibfnamefont {T.}~\bibnamefont
  {Fogarty}}, \bibinfo {author} {\bibfnamefont {L.}~\bibnamefont {Ruks}},
  \bibinfo {author} {\bibfnamefont {J.}~\bibnamefont {Li}}, \ and\ \bibinfo
  {author} {\bibfnamefont {T.}~\bibnamefont {Busch}},\ }\href@noop {}
  {\bibfield  {journal} {\bibinfo  {journal} {arXiv preprint arXiv:1806.08506}\
  } (\bibinfo {year} {2018})}\BibitemShut {NoStop}%
\bibitem [{\citenamefont {Garc\'{\i}a-March}\ \emph
  {et~al.}(2014{\natexlab{b}})\citenamefont {Garc\'{\i}a-March}, \citenamefont
  {Juli\'a-D\'{\i}az}, \citenamefont {Astrakharchik}, \citenamefont {Boronat},\
  and\ \citenamefont {Polls}}]{Garcia-MarchPRA2014}%
  \BibitemOpen
  \bibfield  {author} {\bibinfo {author} {\bibfnamefont {M.~A.}\ \bibnamefont
  {Garc\'{\i}a-March}}, \bibinfo {author} {\bibfnamefont {B.}~\bibnamefont
  {Juli\'a-D\'{\i}az}}, \bibinfo {author} {\bibfnamefont {G.~E.}\ \bibnamefont
  {Astrakharchik}}, \bibinfo {author} {\bibfnamefont {J.}~\bibnamefont
  {Boronat}}, \ and\ \bibinfo {author} {\bibfnamefont {A.}~\bibnamefont
  {Polls}},\ }\href@noop {} {\bibfield  {journal} {\bibinfo  {journal} {Phys.
  Rev. A}\ }\textbf {\bibinfo {volume} {90}},\ \bibinfo {pages} {063605}
  (\bibinfo {year} {2014}{\natexlab{b}})}\BibitemShut {NoStop}%
\bibitem [{\citenamefont {Volosniev}\ \emph {et~al.}(2014)\citenamefont
  {Volosniev}, \citenamefont {Fedorov}, \citenamefont {Jensen}, \citenamefont
  {Zinner},\ and\ \citenamefont {Valiente}}]{Valiente2014Multi}%
  \BibitemOpen
  \bibfield  {author} {\bibinfo {author} {\bibfnamefont {A.~G.}\ \bibnamefont
  {Volosniev}}, \bibinfo {author} {\bibfnamefont {D.~V.}\ \bibnamefont
  {Fedorov}}, \bibinfo {author} {\bibfnamefont {A.~S.}\ \bibnamefont {Jensen}},
  \bibinfo {author} {\bibfnamefont {N.~T.}\ \bibnamefont {Zinner}}, \ and\
  \bibinfo {author} {\bibfnamefont {M.}~\bibnamefont {Valiente}},\ }\href@noop
  {} {\bibfield  {journal} {\bibinfo  {journal} {Few-Body Syst.}\ }\textbf
  {\bibinfo {volume} {55}},\ \bibinfo {pages} {839} (\bibinfo {year}
  {2014})}\BibitemShut {NoStop}%
\bibitem [{\citenamefont {Volosniev}\ \emph {et~al.}(2015)\citenamefont
  {Volosniev}, \citenamefont {Fedorov}, \citenamefont {Jensen},\ and\
  \citenamefont {Zinner}}]{volosniev2015hyperspherical}%
  \BibitemOpen
  \bibfield  {author} {\bibinfo {author} {\bibfnamefont {A.~G.}\ \bibnamefont
  {Volosniev}}, \bibinfo {author} {\bibfnamefont {D.~V.}\ \bibnamefont
  {Fedorov}}, \bibinfo {author} {\bibfnamefont {A.~S.}\ \bibnamefont {Jensen}},
  \ and\ \bibinfo {author} {\bibfnamefont {N.~T.}\ \bibnamefont {Zinner}},\
  }\href@noop {} {\bibfield  {journal} {\bibinfo  {journal} {The European
  Physical Journal Special Topics}\ }\textbf {\bibinfo {volume} {224}},\
  \bibinfo {pages} {585} (\bibinfo {year} {2015})}\BibitemShut {NoStop}%
\bibitem [{\citenamefont {Ko{\'s}cik}(2012)}]{koscik2012quantum}%
  \BibitemOpen
  \bibfield  {author} {\bibinfo {author} {\bibfnamefont {P.}~\bibnamefont
  {Ko{\'s}cik}},\ }\href@noop {} {\bibfield  {journal} {\bibinfo  {journal}
  {Few Body Syst.}\ }\textbf {\bibinfo {volume} {52}},\ \bibinfo {pages} {49}
  (\bibinfo {year} {2012})}\BibitemShut {NoStop}%
\bibitem [{\citenamefont {Zinner}\ and\ \citenamefont
  {Jensen}(2013)}]{zinner2013comparing}%
  \BibitemOpen
  \bibfield  {author} {\bibinfo {author} {\bibfnamefont {N.~T.}\ \bibnamefont
  {Zinner}}\ and\ \bibinfo {author} {\bibfnamefont {A.~S.}\ \bibnamefont
  {Jensen}},\ }\href@noop {} {\bibfield  {journal} {\bibinfo  {journal}
  {Journal of Physics G: Nuclear and Particle Physics}\ }\textbf {\bibinfo
  {volume} {40}},\ \bibinfo {pages} {053101} (\bibinfo {year}
  {2013})}\BibitemShut {NoStop}%
\bibitem [{\citenamefont {D'Amico}\ and\ \citenamefont
  {Rontani}(2015)}]{DAmico2015Pairing}%
  \BibitemOpen
  \bibfield  {author} {\bibinfo {author} {\bibfnamefont {P.}~\bibnamefont
  {D'Amico}}\ and\ \bibinfo {author} {\bibfnamefont {M.}~\bibnamefont
  {Rontani}},\ }\href@noop {} {\bibfield  {journal} {\bibinfo  {journal} {Phys.
  Rev. A}\ }\textbf {\bibinfo {volume} {91}},\ \bibinfo {pages} {043610}
  (\bibinfo {year} {2015})}\BibitemShut {NoStop}%
\bibitem [{\citenamefont {Sowi{\'n}ski}\ \emph {et~al.}(2015)\citenamefont
  {Sowi{\'n}ski}, \citenamefont {Gajda},\ and\ \citenamefont
  {Rz\c{a}{\.z}ewski}}]{Sowinski2015Pairing}%
  \BibitemOpen
  \bibfield  {author} {\bibinfo {author} {\bibfnamefont {T.}~\bibnamefont
  {Sowi{\'n}ski}}, \bibinfo {author} {\bibfnamefont {M.}~\bibnamefont {Gajda}},
  \ and\ \bibinfo {author} {\bibfnamefont {K.}~\bibnamefont
  {Rz\c{a}{\.z}ewski}},\ }\href@noop {} {\bibfield  {journal} {\bibinfo
  {journal} {Europhys. Lett.}\ }\textbf {\bibinfo {volume} {109}},\ \bibinfo
  {pages} {26005} (\bibinfo {year} {2015})}\BibitemShut {NoStop}%
\bibitem [{\citenamefont {Bugnion}\ \emph {et~al.}(2013)\citenamefont
  {Bugnion}, \citenamefont {Lofthouse},\ and\ \citenamefont
  {Conduit}}]{conduit2013fflo}%
  \BibitemOpen
  \bibfield  {author} {\bibinfo {author} {\bibfnamefont {P.~O.}\ \bibnamefont
  {Bugnion}}, \bibinfo {author} {\bibfnamefont {J.~A.}\ \bibnamefont
  {Lofthouse}}, \ and\ \bibinfo {author} {\bibfnamefont {G.~J.}\ \bibnamefont
  {Conduit}},\ }\href {\doibase 10.1103/PhysRevLett.111.045301} {\bibfield
  {journal} {\bibinfo  {journal} {Phys. Rev. Lett.}\ }\textbf {\bibinfo
  {volume} {111}},\ \bibinfo {pages} {045301} (\bibinfo {year}
  {2013})}\BibitemShut {NoStop}%
\bibitem [{\citenamefont {Rammelm{\"u}ller}\ \emph {et~al.}(2016)\citenamefont
  {Rammelm{\"u}ller}, \citenamefont {Porter},\ and\ \citenamefont
  {Drut}}]{rammelmuller2016ground}%
  \BibitemOpen
  \bibfield  {author} {\bibinfo {author} {\bibfnamefont {L.}~\bibnamefont
  {Rammelm{\"u}ller}}, \bibinfo {author} {\bibfnamefont {W.~J.}\ \bibnamefont
  {Porter}}, \ and\ \bibinfo {author} {\bibfnamefont {J.~E.}\ \bibnamefont
  {Drut}},\ }\href@noop {} {\bibfield  {journal} {\bibinfo  {journal} {Physical
  Review A}\ }\textbf {\bibinfo {volume} {93}},\ \bibinfo {pages} {033639}
  (\bibinfo {year} {2016})}\BibitemShut {NoStop}%
\bibitem [{\citenamefont {McKenney}\ \emph {et~al.}(2016)\citenamefont
  {McKenney}, \citenamefont {Shill}, \citenamefont {Porter},\ and\
  \citenamefont {Drut}}]{mckenney2016ground}%
  \BibitemOpen
  \bibfield  {author} {\bibinfo {author} {\bibfnamefont {J.}~\bibnamefont
  {McKenney}}, \bibinfo {author} {\bibfnamefont {C.}~\bibnamefont {Shill}},
  \bibinfo {author} {\bibfnamefont {W.}~\bibnamefont {Porter}}, \ and\ \bibinfo
  {author} {\bibfnamefont {J.}~\bibnamefont {Drut}},\ }\href@noop {} {\bibfield
   {journal} {\bibinfo  {journal} {Journal of Physics B: Atomic, Molecular and
  Optical Physics}\ }\textbf {\bibinfo {volume} {49}},\ \bibinfo {pages}
  {225001} (\bibinfo {year} {2016})}\BibitemShut {NoStop}%
\bibitem [{\citenamefont {Bjerlin}\ \emph {et~al.}(2016)\citenamefont
  {Bjerlin}, \citenamefont {Reimann},\ and\ \citenamefont
  {Bruun}}]{BjerlinReimann2016Higgs}%
  \BibitemOpen
  \bibfield  {author} {\bibinfo {author} {\bibfnamefont {J.}~\bibnamefont
  {Bjerlin}}, \bibinfo {author} {\bibfnamefont {S.~M.}\ \bibnamefont
  {Reimann}}, \ and\ \bibinfo {author} {\bibfnamefont {G.~M.}\ \bibnamefont
  {Bruun}},\ }\href@noop {} {\bibfield  {journal} {\bibinfo  {journal} {Phys.
  Rev. Lett.}\ }\textbf {\bibinfo {volume} {116}},\ \bibinfo {pages} {155302}
  (\bibinfo {year} {2016})}\BibitemShut {NoStop}%
\bibitem [{\citenamefont {Wille}\ \emph {et~al.}(2008)\citenamefont {Wille},
  \citenamefont {Spiegelhalder}, \citenamefont {Kerner}, \citenamefont {Naik},
  \citenamefont {Trenkwalder}, \citenamefont {Hendl}, \citenamefont {Schreck},
  \citenamefont {Grimm}, \citenamefont {Tiecke}, \citenamefont {Walraven},
  \citenamefont {Kokkelmans}, \citenamefont {Tiesinga},\ and\ \citenamefont
  {Julienne}}]{Wille6Li40K}%
  \BibitemOpen
  \bibfield  {author} {\bibinfo {author} {\bibfnamefont {E.}~\bibnamefont
  {Wille}}, \bibinfo {author} {\bibfnamefont {F.~M.}\ \bibnamefont
  {Spiegelhalder}}, \bibinfo {author} {\bibfnamefont {G.}~\bibnamefont
  {Kerner}}, \bibinfo {author} {\bibfnamefont {D.}~\bibnamefont {Naik}},
  \bibinfo {author} {\bibfnamefont {A.}~\bibnamefont {Trenkwalder}}, \bibinfo
  {author} {\bibfnamefont {G.}~\bibnamefont {Hendl}}, \bibinfo {author}
  {\bibfnamefont {F.}~\bibnamefont {Schreck}}, \bibinfo {author} {\bibfnamefont
  {R.}~\bibnamefont {Grimm}}, \bibinfo {author} {\bibfnamefont {T.~G.}\
  \bibnamefont {Tiecke}}, \bibinfo {author} {\bibfnamefont {J.~T.~M.}\
  \bibnamefont {Walraven}}, \bibinfo {author} {\bibfnamefont {S.~J. J. M.~F.}\
  \bibnamefont {Kokkelmans}}, \bibinfo {author} {\bibfnamefont
  {E.}~\bibnamefont {Tiesinga}}, \ and\ \bibinfo {author} {\bibfnamefont
  {P.~S.}\ \bibnamefont {Julienne}},\ }\href@noop {} {\bibfield  {journal}
  {\bibinfo  {journal} {Phys. Rev. Lett.}\ }\textbf {\bibinfo {volume} {100}},\
  \bibinfo {pages} {053201} (\bibinfo {year} {2008})}\BibitemShut {NoStop}%
\bibitem [{\citenamefont {Tiecke}\ \emph {et~al.}(2010)\citenamefont {Tiecke},
  \citenamefont {Goosen}, \citenamefont {Ludewig}, \citenamefont {Gensemer},
  \citenamefont {Kraft}, \citenamefont {Kokkelmans},\ and\ \citenamefont
  {Walraven}}]{tiecke2010Feshbach6Li40K}%
  \BibitemOpen
  \bibfield  {author} {\bibinfo {author} {\bibfnamefont {T.~G.}\ \bibnamefont
  {Tiecke}}, \bibinfo {author} {\bibfnamefont {M.~R.}\ \bibnamefont {Goosen}},
  \bibinfo {author} {\bibfnamefont {A.}~\bibnamefont {Ludewig}}, \bibinfo
  {author} {\bibfnamefont {S.~D.}\ \bibnamefont {Gensemer}}, \bibinfo {author}
  {\bibfnamefont {S.}~\bibnamefont {Kraft}}, \bibinfo {author} {\bibfnamefont
  {S.~J. J. M.~F.}\ \bibnamefont {Kokkelmans}}, \ and\ \bibinfo {author}
  {\bibfnamefont {J.~T.~M.}\ \bibnamefont {Walraven}},\ }\href@noop {}
  {\bibfield  {journal} {\bibinfo  {journal} {Phys. Rev. Lett.}\ }\textbf
  {\bibinfo {volume} {104}},\ \bibinfo {pages} {053202} (\bibinfo {year}
  {2010})}\BibitemShut {NoStop}%
\bibitem [{\citenamefont {Cetina}\ \emph {et~al.}(2016)\citenamefont {Cetina},
  \citenamefont {Jag}, \citenamefont {Lous}, \citenamefont {Fritsche},
  \citenamefont {Walraven}, \citenamefont {Grimm}, \citenamefont {Levinsen},
  \citenamefont {Parish}, \citenamefont {Schmidt}, \citenamefont {Knap} \emph
  {et~al.}}]{cetina2016ultrafast}%
  \BibitemOpen
  \bibfield  {author} {\bibinfo {author} {\bibfnamefont {M.}~\bibnamefont
  {Cetina}}, \bibinfo {author} {\bibfnamefont {M.}~\bibnamefont {Jag}},
  \bibinfo {author} {\bibfnamefont {R.~S.}\ \bibnamefont {Lous}}, \bibinfo
  {author} {\bibfnamefont {I.}~\bibnamefont {Fritsche}}, \bibinfo {author}
  {\bibfnamefont {J.~T.}\ \bibnamefont {Walraven}}, \bibinfo {author}
  {\bibfnamefont {R.}~\bibnamefont {Grimm}}, \bibinfo {author} {\bibfnamefont
  {J.}~\bibnamefont {Levinsen}}, \bibinfo {author} {\bibfnamefont {M.~M.}\
  \bibnamefont {Parish}}, \bibinfo {author} {\bibfnamefont {R.}~\bibnamefont
  {Schmidt}}, \bibinfo {author} {\bibfnamefont {M.}~\bibnamefont {Knap}},
  \emph {et~al.},\ }\href@noop {} {\bibfield  {journal} {\bibinfo  {journal}
  {Science}\ }\textbf {\bibinfo {volume} {354}},\ \bibinfo {pages} {96}
  (\bibinfo {year} {2016})}\BibitemShut {NoStop}%
\bibitem [{\citenamefont {Ravensbergen}\ \emph {et~al.}(2018)\citenamefont
  {Ravensbergen}, \citenamefont {Corre}, \citenamefont {Soave}, \citenamefont
  {Kreyer}, \citenamefont {Kirilov},\ and\ \citenamefont
  {Grimm}}]{Grimm2018DyK}%
  \BibitemOpen
  \bibfield  {author} {\bibinfo {author} {\bibfnamefont {C.}~\bibnamefont
  {Ravensbergen}}, \bibinfo {author} {\bibfnamefont {V.}~\bibnamefont {Corre}},
  \bibinfo {author} {\bibfnamefont {E.}~\bibnamefont {Soave}}, \bibinfo
  {author} {\bibfnamefont {M.}~\bibnamefont {Kreyer}}, \bibinfo {author}
  {\bibfnamefont {E.}~\bibnamefont {Kirilov}}, \ and\ \bibinfo {author}
  {\bibfnamefont {R.}~\bibnamefont {Grimm}},\ }\href {\doibase
  10.1103/PhysRevA.98.063624} {\bibfield  {journal} {\bibinfo  {journal} {Phys.
  Rev. A}\ }\textbf {\bibinfo {volume} {98}},\ \bibinfo {pages} {063624}
  (\bibinfo {year} {2018})}\BibitemShut {NoStop}%
\bibitem [{\citenamefont {Hadzibabic}\ \emph {et~al.}(2002)\citenamefont
  {Hadzibabic}, \citenamefont {Stan}, \citenamefont {Dieckmann}, \citenamefont
  {Gupta}, \citenamefont {Zwierlein}, \citenamefont {G\"orlitz},\ and\
  \citenamefont {Ketterle}}]{Hadzibabic2002LiNa}%
  \BibitemOpen
  \bibfield  {author} {\bibinfo {author} {\bibfnamefont {Z.}~\bibnamefont
  {Hadzibabic}}, \bibinfo {author} {\bibfnamefont {C.~A.}\ \bibnamefont
  {Stan}}, \bibinfo {author} {\bibfnamefont {K.}~\bibnamefont {Dieckmann}},
  \bibinfo {author} {\bibfnamefont {S.}~\bibnamefont {Gupta}}, \bibinfo
  {author} {\bibfnamefont {M.~W.}\ \bibnamefont {Zwierlein}}, \bibinfo {author}
  {\bibfnamefont {A.}~\bibnamefont {G\"orlitz}}, \ and\ \bibinfo {author}
  {\bibfnamefont {W.}~\bibnamefont {Ketterle}},\ }\href {\doibase
  10.1103/PhysRevLett.88.160401} {\bibfield  {journal} {\bibinfo  {journal}
  {Phys. Rev. Lett.}\ }\textbf {\bibinfo {volume} {88}},\ \bibinfo {pages}
  {160401} (\bibinfo {year} {2002})}\BibitemShut {NoStop}%
\bibitem [{\citenamefont {G\"unter}\ \emph {et~al.}(2006)\citenamefont
  {G\"unter}, \citenamefont {St\"oferle}, \citenamefont {Moritz}, \citenamefont
  {K\"ohl},\ and\ \citenamefont {Esslinger}}]{Esslinger2006RbK}%
  \BibitemOpen
  \bibfield  {author} {\bibinfo {author} {\bibfnamefont {K.}~\bibnamefont
  {G\"unter}}, \bibinfo {author} {\bibfnamefont {T.}~\bibnamefont
  {St\"oferle}}, \bibinfo {author} {\bibfnamefont {H.}~\bibnamefont {Moritz}},
  \bibinfo {author} {\bibfnamefont {M.}~\bibnamefont {K\"ohl}}, \ and\ \bibinfo
  {author} {\bibfnamefont {T.}~\bibnamefont {Esslinger}},\ }\href {\doibase
  10.1103/PhysRevLett.96.180402} {\bibfield  {journal} {\bibinfo  {journal}
  {Phys. Rev. Lett.}\ }\textbf {\bibinfo {volume} {96}},\ \bibinfo {pages}
  {180402} (\bibinfo {year} {2006})}\BibitemShut {NoStop}%
\bibitem [{\citenamefont {Best}\ \emph {et~al.}(2009)\citenamefont {Best},
  \citenamefont {Will}, \citenamefont {Schneider}, \citenamefont
  {Hackerm\"uller}, \citenamefont {van Oosten}, \citenamefont {Bloch},\ and\
  \citenamefont {L\"uhmann}}]{Bloch2009RbK}%
  \BibitemOpen
  \bibfield  {author} {\bibinfo {author} {\bibfnamefont {T.}~\bibnamefont
  {Best}}, \bibinfo {author} {\bibfnamefont {S.}~\bibnamefont {Will}}, \bibinfo
  {author} {\bibfnamefont {U.}~\bibnamefont {Schneider}}, \bibinfo {author}
  {\bibfnamefont {L.}~\bibnamefont {Hackerm\"uller}}, \bibinfo {author}
  {\bibfnamefont {D.}~\bibnamefont {van Oosten}}, \bibinfo {author}
  {\bibfnamefont {I.}~\bibnamefont {Bloch}}, \ and\ \bibinfo {author}
  {\bibfnamefont {D.-S.}\ \bibnamefont {L\"uhmann}},\ }\href {\doibase
  10.1103/PhysRevLett.102.030408} {\bibfield  {journal} {\bibinfo  {journal}
  {Phys. Rev. Lett.}\ }\textbf {\bibinfo {volume} {102}},\ \bibinfo {pages}
  {030408} (\bibinfo {year} {2009})}\BibitemShut {NoStop}%
\bibitem [{\citenamefont {Wu}\ \emph {et~al.}(2011)\citenamefont {Wu},
  \citenamefont {Santiago}, \citenamefont {Park}, \citenamefont {Ahmadi},\ and\
  \citenamefont {Zwierlein}}]{Wu2011KKLi}%
  \BibitemOpen
  \bibfield  {author} {\bibinfo {author} {\bibfnamefont {C.-H.}\ \bibnamefont
  {Wu}}, \bibinfo {author} {\bibfnamefont {I.}~\bibnamefont {Santiago}},
  \bibinfo {author} {\bibfnamefont {J.~W.}\ \bibnamefont {Park}}, \bibinfo
  {author} {\bibfnamefont {P.}~\bibnamefont {Ahmadi}}, \ and\ \bibinfo {author}
  {\bibfnamefont {M.~W.}\ \bibnamefont {Zwierlein}},\ }\href {\doibase
  10.1103/PhysRevA.84.011601} {\bibfield  {journal} {\bibinfo  {journal} {Phys.
  Rev. A}\ }\textbf {\bibinfo {volume} {84}},\ \bibinfo {pages} {011601}
  (\bibinfo {year} {2011})}\BibitemShut {NoStop}%
\bibitem [{\citenamefont {Tung}\ \emph {et~al.}(2014)\citenamefont {Tung},
  \citenamefont {Jim\'enez-Garc\'{\i}a}, \citenamefont {Johansen},
  \citenamefont {Parker},\ and\ \citenamefont {Chin}}]{Chin2014CsLi}%
  \BibitemOpen
  \bibfield  {author} {\bibinfo {author} {\bibfnamefont {S.-K.}\ \bibnamefont
  {Tung}}, \bibinfo {author} {\bibfnamefont {K.}~\bibnamefont
  {Jim\'enez-Garc\'{\i}a}}, \bibinfo {author} {\bibfnamefont {J.}~\bibnamefont
  {Johansen}}, \bibinfo {author} {\bibfnamefont {C.~V.}\ \bibnamefont
  {Parker}}, \ and\ \bibinfo {author} {\bibfnamefont {C.}~\bibnamefont
  {Chin}},\ }\href {\doibase 10.1103/PhysRevLett.113.240402} {\bibfield
  {journal} {\bibinfo  {journal} {Phys. Rev. Lett.}\ }\textbf {\bibinfo
  {volume} {113}},\ \bibinfo {pages} {240402} (\bibinfo {year}
  {2014})}\BibitemShut {NoStop}%
\bibitem [{\citenamefont {Lous}\ \emph {et~al.}(2018)\citenamefont {Lous},
  \citenamefont {Fritsche}, \citenamefont {Jag}, \citenamefont {Lehmann},
  \citenamefont {Kirilov}, \citenamefont {Huang},\ and\ \citenamefont
  {Grimm}}]{Grimm2018KLi}%
  \BibitemOpen
  \bibfield  {author} {\bibinfo {author} {\bibfnamefont {R.~S.}\ \bibnamefont
  {Lous}}, \bibinfo {author} {\bibfnamefont {I.}~\bibnamefont {Fritsche}},
  \bibinfo {author} {\bibfnamefont {M.}~\bibnamefont {Jag}}, \bibinfo {author}
  {\bibfnamefont {F.}~\bibnamefont {Lehmann}}, \bibinfo {author} {\bibfnamefont
  {E.}~\bibnamefont {Kirilov}}, \bibinfo {author} {\bibfnamefont
  {B.}~\bibnamefont {Huang}}, \ and\ \bibinfo {author} {\bibfnamefont
  {R.}~\bibnamefont {Grimm}},\ }\href {\doibase 10.1103/PhysRevLett.120.243403}
  {\bibfield  {journal} {\bibinfo  {journal} {Phys. Rev. Lett.}\ }\textbf
  {\bibinfo {volume} {120}},\ \bibinfo {pages} {243403} (\bibinfo {year}
  {2018})}\BibitemShut {NoStop}%
\bibitem [{\citenamefont {Cui}\ and\ \citenamefont
  {Ho}(2013)}]{JasonHo2013PhaseSeparation}%
  \BibitemOpen
  \bibfield  {author} {\bibinfo {author} {\bibfnamefont {X.}~\bibnamefont
  {Cui}}\ and\ \bibinfo {author} {\bibfnamefont {T.~L.}\ \bibnamefont {Ho}},\
  }\href@noop {} {\bibfield  {journal} {\bibinfo  {journal} {Phys. Rev. Lett.}\
  }\textbf {\bibinfo {volume} {110}},\ \bibinfo {pages} {165302} (\bibinfo
  {year} {2013})}\BibitemShut {NoStop}%
\bibitem [{\citenamefont {Loft}\ \emph {et~al.}(2015)\citenamefont {Loft},
  \citenamefont {Dehkharghani}, \citenamefont {Mehta}, \citenamefont
  {Volosniev},\ and\ \citenamefont {Zinner}}]{loft2014variational}%
  \BibitemOpen
  \bibfield  {author} {\bibinfo {author} {\bibfnamefont {N.~J.~S.}\
  \bibnamefont {Loft}}, \bibinfo {author} {\bibfnamefont {A.~S.}\ \bibnamefont
  {Dehkharghani}}, \bibinfo {author} {\bibfnamefont {N.~P.}\ \bibnamefont
  {Mehta}}, \bibinfo {author} {\bibfnamefont {A.~G.}\ \bibnamefont
  {Volosniev}}, \ and\ \bibinfo {author} {\bibfnamefont {N.~T.}\ \bibnamefont
  {Zinner}},\ }\href@noop {} {\bibfield  {journal} {\bibinfo  {journal} {EPJ
  D}\ }\textbf {\bibinfo {volume} {69}},\ \bibinfo {pages} {65} (\bibinfo
  {year} {2015})}\BibitemShut {NoStop}%
\bibitem [{\citenamefont {P{\k e}cak}\ \emph {et~al.}(2016)\citenamefont {P{\k
  e}cak}, \citenamefont {Gajda},\ and\ \citenamefont
  {Sowi\'nski}}]{Pecak2016Separation}%
  \BibitemOpen
  \bibfield  {author} {\bibinfo {author} {\bibfnamefont {D.}~\bibnamefont {P{\k
  e}cak}}, \bibinfo {author} {\bibfnamefont {M.}~\bibnamefont {Gajda}}, \ and\
  \bibinfo {author} {\bibfnamefont {T.}~\bibnamefont {Sowi\'nski}},\
  }\href@noop {} {\bibfield  {journal} {\bibinfo  {journal} {New J. Phys.}\
  }\textbf {\bibinfo {volume} {18}},\ \bibinfo {pages} {013030} (\bibinfo
  {year} {2016})}\BibitemShut {NoStop}%
\bibitem [{\citenamefont {P\ifmmode~\mbox{\k{e}}\else \k{e}\fi{}cak}\ \emph
  {et~al.}(2017)\citenamefont {P\ifmmode~\mbox{\k{e}}\else \k{e}\fi{}cak},
  \citenamefont {Dehkharghani}, \citenamefont {Zinner},\ and\ \citenamefont
  {Sowi\ifmmode~\acute{n}\else \'{n}\fi{}ski}}]{Pecak2017Ansatz}%
  \BibitemOpen
  \bibfield  {author} {\bibinfo {author} {\bibfnamefont {D.}~\bibnamefont
  {P\ifmmode~\mbox{\k{e}}\else \k{e}\fi{}cak}}, \bibinfo {author}
  {\bibfnamefont {A.~S.}\ \bibnamefont {Dehkharghani}}, \bibinfo {author}
  {\bibfnamefont {N.~T.}\ \bibnamefont {Zinner}}, \ and\ \bibinfo {author}
  {\bibfnamefont {T.}~\bibnamefont {Sowi\ifmmode~\acute{n}\else
  \'{n}\fi{}ski}},\ }\href@noop {} {\bibfield  {journal} {\bibinfo  {journal}
  {Phys. Rev. A}\ }\textbf {\bibinfo {volume} {95}},\ \bibinfo {pages} {053632}
  (\bibinfo {year} {2017})}\BibitemShut {NoStop}%
\bibitem [{\citenamefont {Harshman}\ \emph {et~al.}(2017)\citenamefont
  {Harshman}, \citenamefont {Olshanii}, \citenamefont {Dehkharghani},
  \citenamefont {Volosniev}, \citenamefont {Jackson},\ and\ \citenamefont
  {Zinner}}]{harshman2017masses}%
  \BibitemOpen
  \bibfield  {author} {\bibinfo {author} {\bibfnamefont {N.~L.}\ \bibnamefont
  {Harshman}}, \bibinfo {author} {\bibfnamefont {M.}~\bibnamefont {Olshanii}},
  \bibinfo {author} {\bibfnamefont {A.~S.}\ \bibnamefont {Dehkharghani}},
  \bibinfo {author} {\bibfnamefont {A.~G.}\ \bibnamefont {Volosniev}}, \bibinfo
  {author} {\bibfnamefont {S.~G.}\ \bibnamefont {Jackson}}, \ and\ \bibinfo
  {author} {\bibfnamefont {N.~T.}\ \bibnamefont {Zinner}},\ }\href {\doibase
  10.1103/PhysRevX.7.041001} {\bibfield  {journal} {\bibinfo  {journal} {Phys.
  Rev. X}\ }\textbf {\bibinfo {volume} {7}},\ \bibinfo {pages} {041001}
  (\bibinfo {year} {2017})}\BibitemShut {NoStop}%
\bibitem [{\citenamefont {Mistakidis}\ \emph {et~al.}(2018)\citenamefont
  {Mistakidis}, \citenamefont {Katsimiga}, \citenamefont {Koutentakis},\ and\
  \citenamefont {Schmelcher}}]{mistakidis2018repulsive}%
  \BibitemOpen
  \bibfield  {author} {\bibinfo {author} {\bibfnamefont {S.}~\bibnamefont
  {Mistakidis}}, \bibinfo {author} {\bibfnamefont {G.}~\bibnamefont
  {Katsimiga}}, \bibinfo {author} {\bibfnamefont {G.}~\bibnamefont
  {Koutentakis}}, \ and\ \bibinfo {author} {\bibfnamefont {P.}~\bibnamefont
  {Schmelcher}},\ }\href@noop {} {\bibfield  {journal} {\bibinfo  {journal}
  {arXiv preprint arXiv:1808.00040}\ } (\bibinfo {year} {2018})}\BibitemShut
  {NoStop}%
\bibitem [{\citenamefont {Olshanii}(1998)}]{Olshanii1998}%
  \BibitemOpen
  \bibfield  {author} {\bibinfo {author} {\bibfnamefont {M.}~\bibnamefont
  {Olshanii}},\ }\href@noop {} {\bibfield  {journal} {\bibinfo  {journal}
  {Phys. Rev. Lett.}\ }\textbf {\bibinfo {volume} {81}},\ \bibinfo {pages}
  {938} (\bibinfo {year} {1998})}\BibitemShut {NoStop}%
\bibitem [{\citenamefont {Lehoucq}\ \emph {et~al.}(1998)\citenamefont
  {Lehoucq}, \citenamefont {Sorensen},\ and\ \citenamefont
  {Yang}}]{ARPACK1998Sorensen}%
  \BibitemOpen
  \bibfield  {author} {\bibinfo {author} {\bibfnamefont {R.~B.}\ \bibnamefont
  {Lehoucq}}, \bibinfo {author} {\bibfnamefont {D.~C.}\ \bibnamefont
  {Sorensen}}, \ and\ \bibinfo {author} {\bibfnamefont {C.}~\bibnamefont
  {Yang}},\ }\href@noop {} {\emph {\bibinfo {title} {Arpack Users Guide:
  Solution of Large-Scale Eigenvalue Problems With Implicityly Restorted
  Arnoldi Methods}}}\ (\bibinfo  {publisher} {Society for Industrial \& Applied
  Mathematics},\ \bibinfo {address} {Philadelphia},\ \bibinfo {year}
  {1998})\BibitemShut {NoStop}%
\bibitem [{\citenamefont {P{\k{e}}cak}\ \emph {et~al.}(2017)\citenamefont
  {P{\k{e}}cak}, \citenamefont {Gajda},\ and\ \citenamefont
  {Sowi{\'n}ski}}]{Pecak2017com}%
  \BibitemOpen
  \bibfield  {author} {\bibinfo {author} {\bibfnamefont {D.}~\bibnamefont
  {P{\k{e}}cak}}, \bibinfo {author} {\bibfnamefont {M.}~\bibnamefont {Gajda}},
  \ and\ \bibinfo {author} {\bibfnamefont {T.}~\bibnamefont {Sowi{\'n}ski}},\
  }\href@noop {} {\bibfield  {journal} {\bibinfo  {journal} {Few-Body Systems}\
  }\textbf {\bibinfo {volume} {58}},\ \bibinfo {pages} {159} (\bibinfo {year}
  {2017})}\BibitemShut {NoStop}%
\bibitem [{\citenamefont {Altman}\ \emph {et~al.}(2004)\citenamefont {Altman},
  \citenamefont {Demler},\ and\ \citenamefont {Lukin}}]{lukin2004noisCorr}%
  \BibitemOpen
  \bibfield  {author} {\bibinfo {author} {\bibfnamefont {E.}~\bibnamefont
  {Altman}}, \bibinfo {author} {\bibfnamefont {E.}~\bibnamefont {Demler}}, \
  and\ \bibinfo {author} {\bibfnamefont {M.~D.}\ \bibnamefont {Lukin}},\ }\href
  {\doibase 10.1103/PhysRevA.70.013603} {\bibfield  {journal} {\bibinfo
  {journal} {Phys. Rev. A}\ }\textbf {\bibinfo {volume} {70}},\ \bibinfo
  {pages} {013603} (\bibinfo {year} {2004})}\BibitemShut {NoStop}%
\bibitem [{\citenamefont {Mathey}\ \emph {et~al.}(2008)\citenamefont {Mathey},
  \citenamefont {Altman},\ and\ \citenamefont {Vishwanath}}]{Altman20081D}%
  \BibitemOpen
  \bibfield  {author} {\bibinfo {author} {\bibfnamefont {L.}~\bibnamefont
  {Mathey}}, \bibinfo {author} {\bibfnamefont {E.}~\bibnamefont {Altman}}, \
  and\ \bibinfo {author} {\bibfnamefont {A.}~\bibnamefont {Vishwanath}},\
  }\href {\doibase 10.1103/PhysRevLett.100.240401} {\bibfield  {journal}
  {\bibinfo  {journal} {Phys. Rev. Lett.}\ }\textbf {\bibinfo {volume} {100}},\
  \bibinfo {pages} {240401} (\bibinfo {year} {2008})}\BibitemShut {NoStop}%
\bibitem [{\citenamefont {Mathey}\ \emph {et~al.}(2009)\citenamefont {Mathey},
  \citenamefont {Vishwanath},\ and\ \citenamefont {Altman}}]{Altman2009lowDim}%
  \BibitemOpen
  \bibfield  {author} {\bibinfo {author} {\bibfnamefont {L.}~\bibnamefont
  {Mathey}}, \bibinfo {author} {\bibfnamefont {A.}~\bibnamefont {Vishwanath}},
  \ and\ \bibinfo {author} {\bibfnamefont {E.}~\bibnamefont {Altman}},\ }\href
  {\doibase 10.1103/PhysRevA.79.013609} {\bibfield  {journal} {\bibinfo
  {journal} {Phys. Rev. A}\ }\textbf {\bibinfo {volume} {79}},\ \bibinfo
  {pages} {013609} (\bibinfo {year} {2009})}\BibitemShut {NoStop}%
\bibitem [{\citenamefont {Brandt}\ \emph {et~al.}(2017)\citenamefont {Brandt},
  \citenamefont {Yannouleas},\ and\ \citenamefont {Landman}}]{brandt2017two}%
  \BibitemOpen
  \bibfield  {author} {\bibinfo {author} {\bibfnamefont {B.~B.}\ \bibnamefont
  {Brandt}}, \bibinfo {author} {\bibfnamefont {C.}~\bibnamefont {Yannouleas}},
  \ and\ \bibinfo {author} {\bibfnamefont {U.}~\bibnamefont {Landman}},\
  }\href@noop {} {\bibfield  {journal} {\bibinfo  {journal} {Physical Review
  A}\ }\textbf {\bibinfo {volume} {96}},\ \bibinfo {pages} {053632} (\bibinfo
  {year} {2017})}\BibitemShut {NoStop}%
\bibitem [{\citenamefont {Brandt}\ \emph {et~al.}(2018)\citenamefont {Brandt},
  \citenamefont {Yannouleas},\ and\ \citenamefont {Landman}}]{brandt2018HOM}%
  \BibitemOpen
  \bibfield  {author} {\bibinfo {author} {\bibfnamefont {B.~B.}\ \bibnamefont
  {Brandt}}, \bibinfo {author} {\bibfnamefont {C.}~\bibnamefont {Yannouleas}},
  \ and\ \bibinfo {author} {\bibfnamefont {U.}~\bibnamefont {Landman}},\ }\href
  {\doibase 10.1103/PhysRevA.97.053601} {\bibfield  {journal} {\bibinfo
  {journal} {Phys. Rev. A}\ }\textbf {\bibinfo {volume} {97}},\ \bibinfo
  {pages} {053601} (\bibinfo {year} {2018})}\BibitemShut {NoStop}%
\bibitem [{\citenamefont {F{\"o}lling}\ \emph {et~al.}(2005)\citenamefont
  {F{\"o}lling}, \citenamefont {Gerbier}, \citenamefont {Widera}, \citenamefont
  {Mandel}, \citenamefont {Gericke},\ and\ \citenamefont
  {Bloch}}]{folling2005spatial}%
  \BibitemOpen
  \bibfield  {author} {\bibinfo {author} {\bibfnamefont {S.}~\bibnamefont
  {F{\"o}lling}}, \bibinfo {author} {\bibfnamefont {F.}~\bibnamefont
  {Gerbier}}, \bibinfo {author} {\bibfnamefont {A.}~\bibnamefont {Widera}},
  \bibinfo {author} {\bibfnamefont {O.}~\bibnamefont {Mandel}}, \bibinfo
  {author} {\bibfnamefont {T.}~\bibnamefont {Gericke}}, \ and\ \bibinfo
  {author} {\bibfnamefont {I.}~\bibnamefont {Bloch}},\ }\href@noop {}
  {\bibfield  {journal} {\bibinfo  {journal} {Nature}\ }\textbf {\bibinfo
  {volume} {434}},\ \bibinfo {pages} {481} (\bibinfo {year}
  {2005})}\BibitemShut {NoStop}%
\bibitem [{\citenamefont {Deza}\ and\ \citenamefont
  {Deza}(2009)}]{deza2009encyclopedia}%
  \BibitemOpen
  \bibfield  {author} {\bibinfo {author} {\bibfnamefont {M.~M.}\ \bibnamefont
  {Deza}}\ and\ \bibinfo {author} {\bibfnamefont {E.}~\bibnamefont {Deza}},\
  }in\ \href@noop {} {\emph {\bibinfo {booktitle} {Encyclopedia of
  Distances}}}\ (\bibinfo  {publisher} {Springer},\ \bibinfo {year} {2009})\
  pp.\ \bibinfo {pages} {1--583}\BibitemShut {NoStop}%
\bibitem [{\citenamefont {Rachev}\ \emph {et~al.}(2013)\citenamefont {Rachev},
  \citenamefont {Klebanov}, \citenamefont {Stoyanov},\ and\ \citenamefont
  {Fabozzi}}]{rachev2013methods}%
  \BibitemOpen
  \bibfield  {author} {\bibinfo {author} {\bibfnamefont {S.~T.}\ \bibnamefont
  {Rachev}}, \bibinfo {author} {\bibfnamefont {L.}~\bibnamefont {Klebanov}},
  \bibinfo {author} {\bibfnamefont {S.~V.}\ \bibnamefont {Stoyanov}}, \ and\
  \bibinfo {author} {\bibfnamefont {F.}~\bibnamefont {Fabozzi}},\ }\href@noop
  {} {\emph {\bibinfo {title} {The methods of distances in the theory of
  probability and statistics}}}\ (\bibinfo  {publisher} {Springer Science \&
  Business Media},\ \bibinfo {year} {2013})\BibitemShut {NoStop}%
\bibitem [{\citenamefont {Busch}\ \emph {et~al.}(1998)\citenamefont {Busch},
  \citenamefont {Englert}, \citenamefont {Rz\c{a}{\.z}ewski},\ and\
  \citenamefont {Wilkens}}]{Busch1998}%
  \BibitemOpen
  \bibfield  {author} {\bibinfo {author} {\bibfnamefont {T.}~\bibnamefont
  {Busch}}, \bibinfo {author} {\bibfnamefont {B.~G.}\ \bibnamefont {Englert}},
  \bibinfo {author} {\bibfnamefont {K.}~\bibnamefont {Rz\c{a}{\.z}ewski}}, \
  and\ \bibinfo {author} {\bibfnamefont {M.}~\bibnamefont {Wilkens}},\
  }\href@noop {} {\bibfield  {journal} {\bibinfo  {journal} {Found. Phys.}\
  }\textbf {\bibinfo {volume} {28}},\ \bibinfo {pages} {549} (\bibinfo {year}
  {1998})}\BibitemShut {NoStop}%
\bibitem [{\citenamefont {P{\k e}cak}\ and\ \citenamefont
  {Sowi\'nski}(2016)}]{Pecak2016Transition}%
  \BibitemOpen
  \bibfield  {author} {\bibinfo {author} {\bibfnamefont {D.}~\bibnamefont {P{\k
  e}cak}}\ and\ \bibinfo {author} {\bibfnamefont {T.}~\bibnamefont
  {Sowi\'nski}},\ }\href@noop {} {\bibfield  {journal} {\bibinfo  {journal}
  {Phys. Rev. A}\ }\textbf {\bibinfo {volume} {94}},\ \bibinfo {pages} {042118}
  (\bibinfo {year} {2016})}\BibitemShut {NoStop}%
\bibitem [{\citenamefont {von Keyserlingk}\ and\ \citenamefont
  {Conduit}(2011)}]{conduit2011itinerant}%
  \BibitemOpen
  \bibfield  {author} {\bibinfo {author} {\bibfnamefont {C.~W.}\ \bibnamefont
  {von Keyserlingk}}\ and\ \bibinfo {author} {\bibfnamefont {G.~J.}\
  \bibnamefont {Conduit}},\ }\href@noop {} {\bibfield  {journal} {\bibinfo
  {journal} {Phys. Rev. A}\ }\textbf {\bibinfo {volume} {83}},\ \bibinfo
  {pages} {053625} (\bibinfo {year} {2011})}\BibitemShut {NoStop}%
\bibitem [{\citenamefont {Fratini}\ and\ \citenamefont
  {Pilati}(2014)}]{Fratini2014ZeroTemp}%
  \BibitemOpen
  \bibfield  {author} {\bibinfo {author} {\bibfnamefont {E.}~\bibnamefont
  {Fratini}}\ and\ \bibinfo {author} {\bibfnamefont {S.}~\bibnamefont
  {Pilati}},\ }\href@noop {} {\bibfield  {journal} {\bibinfo  {journal} {Phys.
  Rev. A}\ }\textbf {\bibinfo {volume} {90}},\ \bibinfo {pages} {023605}
  (\bibinfo {year} {2014})}\BibitemShut {NoStop}%
\bibitem [{\citenamefont {Garc{\'\i}a-March}\ \emph {et~al.}(2016)\citenamefont
  {Garc{\'\i}a-March}, \citenamefont {Dehkharghani},\ and\ \citenamefont
  {Zinner}}]{garcia2016entanglement}%
  \BibitemOpen
  \bibfield  {author} {\bibinfo {author} {\bibfnamefont {M.}~\bibnamefont
  {Garc{\'\i}a-March}}, \bibinfo {author} {\bibfnamefont {A.~S.}\ \bibnamefont
  {Dehkharghani}}, \ and\ \bibinfo {author} {\bibfnamefont {N.}~\bibnamefont
  {Zinner}},\ }\href@noop {} {\bibfield  {journal} {\bibinfo  {journal}
  {Journal of Physics B: Atomic, Molecular and Optical Physics}\ }\textbf
  {\bibinfo {volume} {49}},\ \bibinfo {pages} {075303} (\bibinfo {year}
  {2016})}\BibitemShut {NoStop}%
\bibitem [{\citenamefont {Girardeau}(1960)}]{Girardeau1960}%
  \BibitemOpen
  \bibfield  {author} {\bibinfo {author} {\bibfnamefont {M.}~\bibnamefont
  {Girardeau}},\ }\href@noop {} {\bibfield  {journal} {\bibinfo  {journal} {J.
  Math. Phys.}\ }\textbf {\bibinfo {volume} {1}},\ \bibinfo {pages} {516}
  (\bibinfo {year} {1960})}\BibitemShut {NoStop}%
\bibitem [{\citenamefont {Paredes}\ \emph {et~al.}(2004)\citenamefont
  {Paredes}, \citenamefont {Widera}, \citenamefont {Murg}, \citenamefont
  {Mandel}, \citenamefont {F{\"o}lling}, \citenamefont {Cirac}, \citenamefont
  {Shlyapnikov}, \citenamefont {H{\"a}nsch},\ and\ \citenamefont
  {Bloch}}]{paredes2004tonks}%
  \BibitemOpen
  \bibfield  {author} {\bibinfo {author} {\bibfnamefont {B.}~\bibnamefont
  {Paredes}}, \bibinfo {author} {\bibfnamefont {A.}~\bibnamefont {Widera}},
  \bibinfo {author} {\bibfnamefont {V.}~\bibnamefont {Murg}}, \bibinfo {author}
  {\bibfnamefont {O.}~\bibnamefont {Mandel}}, \bibinfo {author} {\bibfnamefont
  {S.}~\bibnamefont {F{\"o}lling}}, \bibinfo {author} {\bibfnamefont
  {I.}~\bibnamefont {Cirac}}, \bibinfo {author} {\bibfnamefont {G.~V.}\
  \bibnamefont {Shlyapnikov}}, \bibinfo {author} {\bibfnamefont {T.~W.}\
  \bibnamefont {H{\"a}nsch}}, \ and\ \bibinfo {author} {\bibfnamefont
  {I.}~\bibnamefont {Bloch}},\ }\href@noop {} {\bibfield  {journal} {\bibinfo
  {journal} {Nature}\ }\textbf {\bibinfo {volume} {429}},\ \bibinfo {pages}
  {277} (\bibinfo {year} {2004})}\BibitemShut {NoStop}%
\bibitem [{\citenamefont {Kinoshita}\ \emph {et~al.}(2004)\citenamefont
  {Kinoshita}, \citenamefont {Wenger},\ and\ \citenamefont
  {Weiss}}]{kinoshita2004observation}%
  \BibitemOpen
  \bibfield  {author} {\bibinfo {author} {\bibfnamefont {T.}~\bibnamefont
  {Kinoshita}}, \bibinfo {author} {\bibfnamefont {T.}~\bibnamefont {Wenger}}, \
  and\ \bibinfo {author} {\bibfnamefont {D.~S.}\ \bibnamefont {Weiss}},\
  }\href@noop {} {\bibfield  {journal} {\bibinfo  {journal} {Science}\ }\textbf
  {\bibinfo {volume} {305}},\ \bibinfo {pages} {1125} (\bibinfo {year}
  {2004})}\BibitemShut {NoStop}%
\bibitem [{\citenamefont {Stewart}(1984)}]{HFsystems1984}%
  \BibitemOpen
  \bibfield  {author} {\bibinfo {author} {\bibfnamefont {G.~R.}\ \bibnamefont
  {Stewart}},\ }\href {\doibase 10.1103/RevModPhys.56.755} {\bibfield
  {journal} {\bibinfo  {journal} {Rev. Mod. Phys.}\ }\textbf {\bibinfo {volume}
  {56}},\ \bibinfo {pages} {755} (\bibinfo {year} {1984})}\BibitemShut
  {NoStop}%
\bibitem [{\citenamefont {Matsuda}\ and\ \citenamefont
  {Shimahara}(2007)}]{matsuda2007fulde}%
  \BibitemOpen
  \bibfield  {author} {\bibinfo {author} {\bibfnamefont {Y.}~\bibnamefont
  {Matsuda}}\ and\ \bibinfo {author} {\bibfnamefont {H.}~\bibnamefont
  {Shimahara}},\ }\href@noop {} {\bibfield  {journal} {\bibinfo  {journal}
  {Journal of the Physical Society of Japan}\ }\textbf {\bibinfo {volume}
  {76}},\ \bibinfo {pages} {051005} (\bibinfo {year} {2007})}\BibitemShut
  {NoStop}%
\bibitem [{\citenamefont {Kinnunen}\ \emph {et~al.}(2018)\citenamefont
  {Kinnunen}, \citenamefont {Baarsma}, \citenamefont {Martikainen},\ and\
  \citenamefont {T{\"o}rm{\"a}}}]{kinnunen2018fulde}%
  \BibitemOpen
  \bibfield  {author} {\bibinfo {author} {\bibfnamefont {J.~J.}\ \bibnamefont
  {Kinnunen}}, \bibinfo {author} {\bibfnamefont {J.~E.}\ \bibnamefont
  {Baarsma}}, \bibinfo {author} {\bibfnamefont {J.-P.}\ \bibnamefont
  {Martikainen}}, \ and\ \bibinfo {author} {\bibfnamefont {P.}~\bibnamefont
  {T{\"o}rm{\"a}}},\ }\href@noop {} {\bibfield  {journal} {\bibinfo  {journal}
  {Reports on Progress in Physics}\ }\textbf {\bibinfo {volume} {81}},\
  \bibinfo {pages} {046401} (\bibinfo {year} {2018})}\BibitemShut {NoStop}%
\end{thebibliography}%
\end{document}